\documentclass[journal]{IEEEtran}

\usepackage{amsmath, amssymb, amsthm}

\usepackage{algorithm}
\usepackage{algpseudocode}

\usepackage{graphicx}
\usepackage{subcaption}
\usepackage{tabularx}
\usepackage{amssymb} 

\usepackage{setspace}
\usepackage{multicol}

\usepackage[font=small,labelfont=bf]{caption}

\usepackage{listings}
\usepackage{verbatim}
\usepackage{fancyvrb}

\usepackage{xcolor}
\usepackage{booktabs}
\usepackage{algpseudocode}

\usepackage[normalem]{ulem}

\usepackage[utf8]{inputenc}

\DeclareUnicodeCharacter{221E}{\ensuremath{\infty}}

\usepackage[hidelinks]{hyperref}
\hypersetup{
    colorlinks=true,
    linkcolor=blue,
    citecolor=blue,
    urlcolor=blue
}

\begin{document}

\title{Tri-LLM Cooperative Federated Zero-Shot Intrusion Detection with Semantic Disagreement and Trust-Aware Aggregation}

\author{
Saeid~Jamshidi,
Omar~Abdul-Wahab,
Foutse~Khomh,
and Kawser~Wazed~Nafi
\thanks{Corresponding author: Saeid Jamshidi (\texttt{saeid.jamshidi@polymtl.ca}).}%
\thanks{Kawser~Wazed~Nafi, and Foutse Khomh are with the SWAT Laboratory, Polytechnique Montréal, Québec, Canada.}%
\thanks{Saeid Jamshidi, Omar Abdul-Wahab, is with the Department of Computer and Software Engineering, Polytechnique Montréal, Québec, Canada.}%
}

\maketitle

\begin{abstract}
Federated learning (FL) has emerged as an effective paradigm for privacy-preserving and distributed Intrusion Detection Systems (IDS) in cyber-physical and Internet of Things (IoT) networks, where centralized data aggregation is often infeasible due to privacy and bandwidth constraints. Despite its advantages, most existing FL-based IDS operate under closed-set learning assumptions and lack principled mechanisms such as calibrated uncertainty estimation, semantic generalization beyond fixed label spaces, and explicit modeling of epistemic ambiguity under zero-day attack conditions. Furthermore, robustness to heterogeneous and unreliable clients remains a persistent challenge in practical deployments.
This paper introduces a semantics-driven federated IDS framework that integrates language-derived semantic supervision into the federated optimization process, enabling open-set and zero-shot intrusion reasoning for previously unseen attack behaviors. The proposed approach constructs semantic attack prototypes using a Tri-LLM ensemble comprising GPT-4o, DeepSeek-V3, and LLaMA-3-8B, thereby aligning distributed telemetry features with high-level attack concepts in a shared embedding space. Inter-LLM semantic disagreement is explicitly modeled as an epistemic uncertainty signal for zero-day risk estimation, while a trust-aware aggregation mechanism dynamically weights client updates based on reliability to enhance robustness under non-identically distributed data and adversarial conditions. Experimental results demonstrate stable semantic alignment across heterogeneous clients and consistent convergence over federated rounds. The framework achieves over 80\% zero-shot detection accuracy on previously unseen attack patterns and improves zero-day discrimination by more than 10\% compared to similarity-only baselines, while maintaining low aggregation instability in the presence of unreliable and compromised participants.
\end{abstract}


\section{Introduction}
\label{intro}
Modern cyber-physical infrastructures, including industrial Internet of Things (IIoT) systems, smart grids, and distributed edge monitoring platforms, operate under strict privacy, regulatory, and bandwidth constraints that prevent centralized collection of large-scale telemetry data \cite{reyes2025cybersecurity,hossain2024implementation}. Consequently, intrusion detection models must be trained collaboratively across distributed environments without pooling raw data. Federated learning (FL) directly addresses this requirement by enabling decentralized optimization while keeping data local.\\
In parallel, the threat landscape has shifted toward increasingly adaptive and previously unseen attack behaviors, including polymorphic botnets, stealthy lateral movement, and protocol-level mimicry that evade static signatures and fixed taxonomies \cite{chowdhury2025next,el2025sfarp}. This evolution introduces a second challenge: detection systems must operate under open-set conditions, where novel attack categories may appear during deployment without labeled examples. Traditional closed-set classifiers are ill-suited to such environments, as they tend to produce overconfident predictions under distributional shift.\\
FL-based IDS research has primarily focused on addressing the first challenge, distributed privacy-preserving learning. Early works demonstrated the feasibility of collaborative IDS training across edge devices \cite{man2021fed_iot}, followed by fog-assisted architectures and resource-aware coordination strategies \cite{saha2021fogfl}. Ensemble and multi-view FL approaches have further improved robustness under heterogeneous data distributions \cite{attota2021ensemble_fl_ids}. The surveys highlight persistent challenges, including non-identically distributed (non-IID) drift, label scarcity, communication overhead, and vulnerability to poisoning and unreliable clients \cite{agrawal2022review_fl_ids,khraisat2024fl_survey,zhang2025survey_fl_ids,makris2025survey_fl_ids}. However, most existing FL-IDS frameworks remain fundamentally classification-driven and assume fixed label spaces. Recent efforts in semi-supervised and self-supervised IDS attempt to mitigate label scarcity by leveraging representation learning and anomaly-centric objectives \cite{zhao2022ssfl_iiot,fedmse2025}. While such methods reduce dependence on annotated data, they still rely on statistical similarity within observed distributions and do not explicitly model semantic generalization and calibrated epistemic uncertainty for unseen attack categories. In other words, self-supervised learning improves representation quality but does not inherently provide concept-level reasoning and open-set attribution mechanisms.\\
Large Language Models (LLMs) have recently demonstrated strong capabilities in cybersecurity tasks that require semantic abstraction, contextual reasoning, and interpretation under incomplete labels \cite{xu2024large,ali2025beyond}. Surveys report rapid progress in LLM-driven threat analysis, while also highlighting reliability limits, challenges in uncertainty handling, and safety risks \cite{chen2024llm_threat_survey,das2024llm_secpriv_survey,acm2024llm_cyber_slr}. Importantly, LLMs encode high-level semantic relationships between attacker intent, tactics, and observable behaviors—knowledge that is difficult to capture using purely statistical feature learning. Nevertheless, integrating such semantic knowledge into federated IDS pipelines remains largely unexplored. Beyond semantic generalization, robustness under heterogeneous and potentially compromised clients remains a fundamental concern in federated environments. Advanced aggregation and coordination strategies have been proposed to mitigate non-IID effects and improve convergence stability \cite{olanrewaju2025fl_ids_csa,jianping2024attention_gnn_fl,song2024fedacnn}. However, most approaches focus on optimizing statistical model aggregation rather than integrating uncertainty-aware reasoning layers into the detection pipeline. To address these limitations, we propose a semantics-driven federated IDS framework that couples federated optimization with language-derived semantic supervision. The key idea is to learn a global projection that maps low-level telemetry features into a shared semantic embedding space, where attack hypotheses are represented as language-derived prototypes. These prototypes are constructed using a Tri-LLM ensemble composed of GPT-4o, DeepSeek-V3, and LLaMA-3-8B. Importantly, the LLMs are not federated; they are fine-tuned and trained within the FL loop. They serve solely as external semantic encoders, generating attack concept embeddings from textual descriptions. Only the lightweight projection model is optimized collaboratively across clients via FL. To enhance robustness and reduce sensitivity to individual model bias, we explicitly model inter-LLM semantic disagreement as an epistemic uncertainty signal. This enables calibrated zero-day risk estimation rather than forced closed-set classification. Furthermore, we introduce a trust-aware aggregation mechanism that dynamically weights client updates based on the semantic alignment loss, while monitoring trust entropy as a stability diagnostic. This reduces the impact of unreliable, drifting, and partially compromised participants in heterogeneous deployments. The main contributions of this work are:

\begin{itemize}
    \item \textbf{Tri-LLM semantic supervision for FL-IDS:} We introduce a federated IDS paradigm that integrates language-derived semantic prototypes into a shared embedding space, enabling zero-shot attribution under open-set conditions.
    \item \textbf{Disagreement-aware zero-day risk modeling:} We define an inter-LLM semantic disagreement metric and combine it with similarity-based confidence to produce a calibrated zero-day risk score.
    \item \textbf{Trust-aware federated aggregation under heterogeneity:} We propose a loss-driven trust weighting mechanism with entropy-based monitoring to attenuate unreliable or drifting client updates, improving robustness under non-IID and adversarial environments.
\end{itemize}

The remainder of this paper is organized as follows. Section~\ref{sec:related_work} reviews the related literature on federated intrusion detection and LLM-driven cybersecurity. Section~\ref{subsec:threatmodel} formalizes the adversarial assumptions and security model considered in this work. Section~\ref{sec:proposed} presents the proposed Tri-LLM cooperative federated framework, including semantic prototype construction, trust-aware aggregation, and zero-shot inference mechanisms. Section~\ref{sec:results} reports the experimental findings and provides a comprehensive evaluation of semantic alignment, disagreement behavior, trust dynamics, and zero-shot detection performance. Section~\ref{sec:comparative_baselines} offers comparative analysis with federated IDS baselines and representative frameworks. Section~\ref{Discussion} discusses the broader implications of the findings. Section~\ref{Limitations} outlines current limitations, and Section~\ref{sec:future_work} presents future research directions. Section~\ref{Conclusion} concludes the paper.

\section{Related Work}
\label{sec:related_work}
This section reviews recent advances in FL–based IDS and highlights how prior work has addressed privacy preservation, robustness under heterogeneous clients, and limited supervision. We further discuss emerging efforts at incorporating semantic reasoning and explain why existing approaches remain constrained in open-world and zero-day threat settings.

\subsection{FL for IDS in IoT/IIoT}
FL has become a practical alternative to centralized IDS training when traffic logs and telemetry cannot be pooled due to privacy and governance constraints. A representative early design is FedACNN, which demonstrated that convolutional models can be trained across edge-assisted IoT clients while keeping data local, and still achieve competitive detection performance on heterogeneous devices \cite{man2021fedacnn}. Beyond single-model training, multi-view and ensemble-style FL-IDS approaches were introduced to enhance robustness to feature heterogeneity and attack diversity. For example, MV-FLID demonstrated that combining multiple “views” of IoT data under FL can improve detection stability and classification accuracy \cite{pouriyeh2021mvflid}.

As the area matured, survey papers consolidated the field and made recurring limitations explicit. Agrawal et al.\ summarized core FL-IDS concepts and highlighted recurring operational barriers such as non-identically distributed drift, label scarcity, and communication overhead \cite{agrawal2022comcom}. Khraisat et al.\ provided a broader taxonomy of FL-IDS architectures and aggregation strategies, emphasizing security threats (e.g., poisoning) and practical deployment challenges \cite{khraisat2024acmsurv}. More recently, Zhang et al.\ reviewed FL-IDS from an intrusion-detection lifecycle viewpoint (data, modeling, aggregation, evaluation), arguing that robustness and evaluation realism remain under-addressed \cite{zhang2025jpdc}. Makris et al.\ delivered a comprehensive synthesis of federated IDS techniques and systematically compared aggregation and threat assumptions, positioning robustness as a first-class design constraint \cite{makris2025cosrev}.

\subsection{Robust Aggregation and Poisoning-Aware FL-IDS}
A consistent theme in FL-IDS is that “privacy-preserving training” does not automatically imply “robust training.” Poisoning (including label flipping and malicious client updates) can significantly degrade global IDS quality if the server aggregates data naively. Yang et al.\ studied poisoning in FL-based IoT IDS and proposed server-side scoring/filtering mechanisms to mitigate the impact of unreliable participants during aggregation \cite{yang2023dependablefl}. supportive to FL-specific threats, adversarial machine learning against NIDS has also been surveyed in depth: He et al.\ provide a structured view of adversarial objectives, attack surfaces, and defenses, clarifying why IDS settings are particularly exposed to realistic, replayable packet-level manipulations \cite{he2023comst}.

\subsection{Semi-Supervised and Practical Fog/Industrial Deployments}
Because operational security data is rarely fully labeled, recent FL-IDS work increasingly moves toward semi-supervised or weakly supervised regimes. FedMSE proposed a semi-supervised federated framework for IoT IDS that combines representation learning with anomaly-centric decision logic, introducing an aggregation rule that favors higher-quality local models \cite{beuran2025fedmse}. In parallel, fog/industrial settings motivate designs that explicitly trade off detection accuracy against latency and resource constraints. For example, FFL-IDS presented a fog-enabled FL-IDS designed to counter industrial jamming/spoofing scenarios while maintaining training feasibility in constrained environments \cite{rehman2025fflids}. Empirical studies have also started to quantify deployment feasibility: Albanbay et al.\ investigated which local models and data volumes are practical for IoT devices in FL-based IDS deployments, shifting attention from accuracy alone toward “can this actually run on-device?” \cite{albanbay2025iotflids}.

\subsection{LLM-Driven Cyber Semantics and the Gap to Federated IDS}
LLMs are increasingly used in cybersecurity because they can capture semantic structure (tactics, intent, context) that is difficult to encode using purely statistical feature learning. Chen et al.\ surveyed how LLMs are being applied to cyber threat detection tasks and outlined where LLMs add value (e.g., abstraction, interpretability) versus where they remain brittle \cite{chen2024llmthreat}. A broader systematic view is provided by Xu et al., who reviewed LLM4Security research at scale and categorized common application patterns and limitations \cite{xu2024llm4security}. At the same time, the security and privacy risks of LLMs are well-documented: Das et al.\ surveyed attack classes such as jailbreaking, data poisoning, and privacy leakage, emphasizing the need for reliability and uncertainty handling \cite{das2024llmsecpriv}. Recent work also highlights how quickly this space is expanding and why integration patterns matter: Zhang et al.\ reviewed “LLMs meet cybersecurity” and stressed that evaluation rigor and safety constraints remain uneven across tasks \cite{zhang2025llmcyber}.\\

The literature review indicates that FL has become an effective paradigm for privacy-preserving IDS across distributed, heterogeneous environments, with demonstrated gains in scalability and data governance. However, most existing FL-IDS solutions continue to rely on fixed label spaces and purely statistical feature learning, which limits their ability to generalize beyond known attack classes. Recent efforts toward robustness and semi-supervision mitigate some practical challenges but do not explicitly address semantic ambiguity or epistemic uncertainty under open-world conditions. In parallel, LLM-driven security research shows strong potential for capturing high-level attack intent and contextual meaning, yet remains largely disconnected from federated detection pipelines. To bridge this divide, our work integrates language-derived semantic knowledge into the FL process, enabling concept-level intrusion reasoning without centralized data access. By jointly modeling semantic similarity, inter-model disagreement, and client trust, the proposed framework advances the design of federated IDS toward reliable zero-shot and zero-day detection.

\section{Threat Model and Security Analysis}
\label{subsec:threatmodel}
This section formalizes the threat model considered in the proposed Tri-LLM Cooperative Federated Zero-Shot IDS framework and analyzes its security properties under realistic adversarial assumptions. In contrast to purely descriptive threat models, we explicitly characterize adversarial objectives and system defenses using mathematical constructs aligned with the proposed semantic and FL formulation.

\subsubsection{System and Adversary Model}
We consider a distributed cyber-physical environment composed of $N$ edge clients participating in an FL process coordinated by a central server. Each client observes local network traffic and system telemetry and maintains a local semantic projection model. Raw traffic data and extracted features remain local, while only model parameters and scalar statistics are exchanged. The adversary is assumed to be adaptive and capable of generating both known and previously unseen attack behaviors. Let $\mathbf{x} \in \mathbb{R}^d$ denote an observed traffic instance and $\hat{\mathbf{z}} = f(\mathbf{x}; \mathbf{W}_g)$ its projection into the global semantic space. The adversary may control a subset of clients and network entities but does not have direct access to the LLMs' internal parameters or prompts used to construct semantic knowledge.

\subsubsection{Data-Level Evasion Threats}
At the data level, the adversary seeks to evade detection by crafting malicious inputs whose semantic embeddings closely resemble those of benign behavior. Formally, an evasion-oriented adversary attempts to solve
\begin{equation}
\min_{\mathbf{x}'} \; \left\| f(\mathbf{x}'; \mathbf{W}_g) - \mathbf{z}_{\text{benign}} \right\|_2,
\end{equation}
subject to operational constraints that preserve the malicious effect of $\mathbf{x}'$. In contrast to signature-based detectors, the proposed framework operates in a semantic embedding space aligned with high-level attack concepts. As a result, superficial perturbations that preserve malicious intent are unlikely to significantly reduce semantic distance, limiting the effectiveness of low-level evasion strategies.

\subsubsection{Federated Model Poisoning and Client Reliability}
At the FL level, a compromised client $i$ may submit a manipulated update $\mathbf{W}_i^{\mathrm{adv}}$ in an attempt to bias the global model. The impact of such poisoning attacks can be expressed as the deviation induced in the aggregated model:
\begin{equation}
\Delta \mathbf{W}_g = \sum_{i=1}^{N} \alpha_i (\mathbf{W}_i - \mathbf{W}_g),
\end{equation}
where $\alpha_i$ denotes the aggregation weight of client $i$. In the proposed framework, aggregation weights are derived from a trust score inversely proportional to the local semantic alignment loss,
\begin{equation}
\tau_i = \frac{1}{\mathcal{L}_i + \epsilon}, \quad
\alpha_i = \frac{\tau_i}{\sum_{j=1}^{N} \tau_j}.
\end{equation}
Clients whose updates result in significant semantic deviations incur higher losses and consequently receive lower aggregation weights. This mechanism bounds the impact of poisoned updates unless a majority of clients are simultaneously compromised.

\subsubsection{Semantic Uncertainty and Inter-LLM Disagreement}
At the semantic level, uncertainty arises from the variability in LLM interpretations. Let $\mathbf{z}_a^{(m)}$ denote the embedding generated for attack concept $a$ by the $m$-th LLM, where $m \in \{\text{GPT-4o}, \text{DeepSeek-V3}, \text{LLaMA-3-8B}\}$. Semantic disagreement is formalized as the average pairwise dispersion
\begin{equation}
D_a = \frac{1}{M(M-1)} \sum_{m \neq n} \left\| \mathbf{z}_a^{(m)} - \mathbf{z}_a^{(n)} \right\|_2,
\end{equation}
with $M=3$ in the Tri-LLM setting. Rather than treating disagreement as noise, the framework explicitly incorporates $D_a$ as an epistemic uncertainty signal. Elevated disagreement indicates reduced semantic consensus and lowers attribution confidence during inference.

\subsubsection{Zero-Day and Open-World Threats}
Zero-day attacks are characterized by the absence of labeled samples and predefined signatures. Let $\mathbf{z}_a$ denote the fused semantic prototype of attack $a$. Zero-shot attribution relies on semantic similarity
\begin{equation}
\hat{a} = \arg\max_{a \in \mathcal{A}} \cos(\hat{\mathbf{z}}, \mathbf{z}_a),
\end{equation}
while uncertainty is quantified through a zero-day risk score
\begin{equation}
\mathrm{ZDS}(\mathbf{x}) =
\lambda D_{\hat{a}} +
(1-\lambda)\left(1 - \cos(\hat{\mathbf{z}}, \mathbf{z}_{\hat{a}})\right).
\end{equation}
This formulation ensures that samples exhibiting low semantic similarity do not lead to overconfident classification of novel behaviors.

\section{Proposed Method}
\label{sec:proposed}
\begin{figure*}[t]
    \centering
    \includegraphics[width=0.8\linewidth]{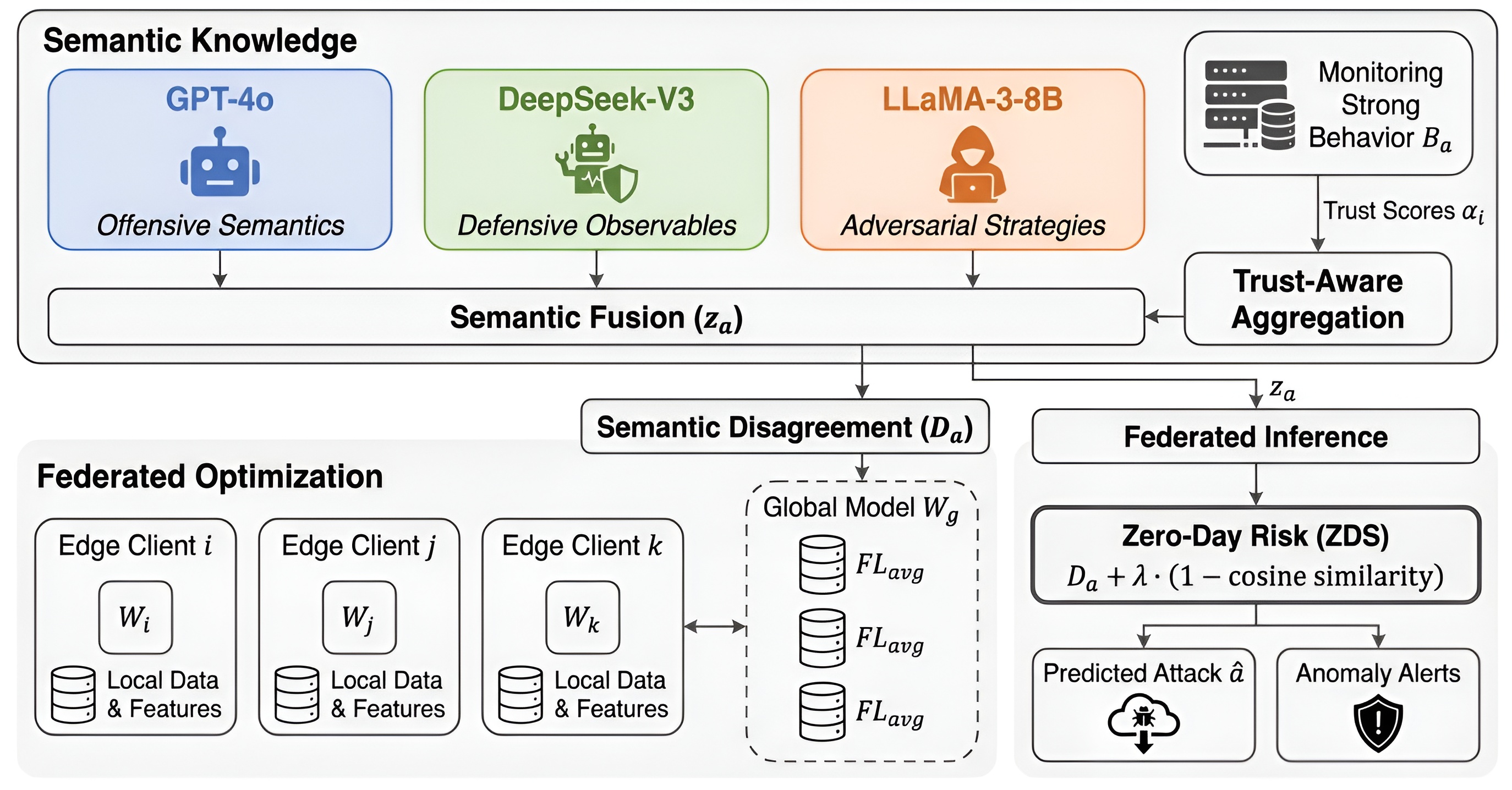}
    \caption{Overview of the proposed Tri-LLM cooperative federated zero-shot IDS framework.}
    \label{fig:overview}
\end{figure*}

This section presents a cooperative, semantics-driven federated IDS framework designed to operate under open-world and zero-day conditions. An overview of the proposed architecture is shown in Figure. \ref{fig:overview}. The framework combines language-derived semantic knowledge, trust-aware federated optimization, and uncertainty-aware inference within a unified semantic embedding space.\\
At a high level, semantic knowledge about attack behaviors is constructed using multiple LLMs and fused into shared semantic prototypes. These prototypes serve as concept-level supervision that aligns distributed client representations without requiring labeled attack samples. FL is employed to optimize client-side semantic projection models under privacy constraints, while a trust-aware aggregation mechanism dynamically controls the impact of heterogeneous, potentially unreliable participants. During inference, projected observations are matched against semantic attack prototypes, and zero-day risk is estimated by jointly considering semantic similarity and inter-model disagreement. Moreover, this design enables robust IDS and calibrated uncertainty estimation in distributed cyber-physical and IoT networks where attack labels are incomplete or unavailable.

\subsection{System Model and Problem Setting}
We consider a distributed environment composed of $N$ edge clients, including IoT gateways, industrial controllers, and monitoring nodes, deployed across heterogeneous administrative domains. Each client continuously observes local network traffic and system telemetry generated by benign operations and adversarial activities. Due to privacy, regulatory, and bandwidth constraints, raw traffic data and extracted features are never shared with clients and the coordinating server. The coordinating server manages the FL process and aggregates model updates. It is assumed to be honest but curious, and to follow the protocol correctly while remaining unable to access raw data. The adversary is assumed to be adaptive and capable of launching novel and zero-day attacks whose signatures and labels are absent from historical training data. Consequently, the detection mechanism must rely on semantic generalization rather than closed-set pattern matching.

\subsection{Local Feature Representation and Semantic Projection}
At each client $i$, raw observations are transformed into a feature vector
\begin{equation}
\mathbf{x}_{i,t} \in \mathbb{R}^{d},
\end{equation}
where $t$ denotes the local time index and $d$ is the feature dimensionality. The feature vector aggregates protocol-level statistics, flow-level summaries, temporal dynamics, and host-level indicators. Feature extraction is deterministic and consistent across all clients to ensure semantic compatibility during federated optimization.
To bridge the gap between low-level features and high-level attack concepts, each client maintains a semantic projection model parameterized by $\mathbf{W}_i \in \mathbb{R}^{k \times d}$. This model maps operational features into a shared semantic embedding space:
\begin{equation}
\hat{\mathbf{z}}_{i,t} = f(\mathbf{x}_{i,t}; \mathbf{W}_i).
\end{equation}
The mapping function $f(\cdot)$ is instantiated as a linear projection $\hat{\mathbf{z}}_{i,t} = \mathbf{W}_i \mathbf{x}_{i,t}$ to enable efficient training on resource-constrained devices. The semantic space is shared across all clients and aligned with language-derived attack representations.

\subsection{Tri-LLM Semantic Knowledge Construction}
To enable zero-shot semantic supervision without relying on labeled attack samples, the framework constructs a semantic knowledge base using three supportive LLMs: GPT-4o, DeepSeek-V3, and LLaMA-3-8B. These models are intentionally selected for their architectural diversity and distinct inference characteristics.
GPT-4o provides contextual reasoning and balanced semantic abstraction, making it suitable for capturing attacker intent and high-level threat descriptions. DeepSeek-V3 is employed for its efficient decoding and relatively low-variance semantic representations, which stabilize the semantic space. LLaMA-3-8B contributes higher expressiveness and sensitivity to fine-grained contextual variations, enabling richer modeling of adversarial execution strategies.
Let $\mathcal{A}$ denote the universe of attack categories, including both known and potentially unseen attacks. For each attack category $a \in \mathcal{A}$, three textual descriptions are generated, each emphasizing a different semantic perspective: (i) an offensive perspective describing attacker goals and tactics, (ii) a defensive perspective focusing on observable network and system symptoms, and (iii) an adversarial strategy perspective capturing execution patterns and evasion behavior. Each description is processed by a distinct LLM, yielding three semantic embeddings:
\begin{equation}
\begin{aligned}
\mathbf{z}_a^{(1)} &= \phi_{\text{GPT-4o}}(\mathbf{s}_a^{\text{off}}), \\
\mathbf{z}_a^{(2)} &= \phi_{\text{DeepSeek-V3}}(\mathbf{s}_a^{\text{def}}), \\
\mathbf{z}_a^{(3)} &= \phi_{\text{LLaMA-3-8B}}(\mathbf{s}_a^{\text{adv}}).
\end{aligned}
\end{equation}

\subsection{Semantic Fusion and Disagreement Modeling}
A unified semantic prototype for each attack category is obtained by averaging the three LLM embeddings:
\begin{equation}
\mathbf{z}_a = \frac{1}{3} \sum_{j=1}^{3} \mathbf{z}_a^{(j)}.
\end{equation}
Although fusion produces a single semantic target for federated training, the diversity among LLM outputs provides valuable epistemic information. Rather than enforcing hard consensus, the proposed framework explicitly models semantic disagreement as an uncertainty signal.
Semantic disagreement for attack category $a$ is quantified as the average pairwise distance among the three embeddings:
\begin{equation}
D_a = \frac{1}{3(3-1)} \sum_{i \neq j} \left\| \mathbf{z}_a^{(i)} - \mathbf{z}_a^{(j)} \right\|_2.
\end{equation}
Higher values of $D_a$ indicate reduced inter-model agreement and increased uncertainty in the semantic characterization of the attack.

\subsection{Federated Training Objective}
During each federated round, client $i$ optimizes its local semantic projection model by minimizing a semantic alignment loss:
\begin{equation}
\mathcal{L}_i =
\frac{1}{|\mathcal{D}_i|} \sum_{(\mathbf{x}, a) \in \mathcal{D}_i}
\left\| f(\mathbf{x}; \mathbf{W}_i) - \mathbf{z}_a \right\|_2^2.
\end{equation}
This objective aligns feature-derived embeddings with language-derived semantic prototypes, enabling the model to generalize beyond explicitly labeled classes.

\subsection{Trust-Aware Federated Aggregation}
After local optimization, each client $i$ transmits its updated projection parameters $\mathbf{W}_i$ along with the corresponding semantic alignment loss $\mathcal{L}_i$. The server assigns a reliability-based trust score that reflects how well the client’s local model aligns with the shared semantic prototype space.
We define the instantaneous trust score as:
\begin{equation}
\tau_i^{(t)} = \frac{1}{\mathcal{L}_i^{(t)} + \epsilon},
\end{equation}
where $\epsilon > 0$ is a stabilizing constant to prevent division by zero. This formulation is motivated by interpreting the semantic alignment loss as an inverse proxy for representation reliability: clients whose local embeddings are well aligned with global semantic prototypes produce lower loss values and are therefore considered more trustworthy contributors. Unlike identity-based or behavioral trust models, this formulation derives trust directly from the quality of task-consistent optimization. To avoid reacting excessively to transient fluctuations, we introduce temporal smoothing using exponential averaging:
\begin{equation}
u_i^{(t)} = \gamma u_i^{(t-1)} + (1-\gamma)\tau_i^{(t)},
\end{equation}
where $\gamma \in [0,1)$ controls the memory of the trust process. This update rule ensures that trust evolves gradually over federated rounds, integrating both historical consistency and recent behavior. Consequently, clients must demonstrate sustained alignment quality to maintain high trust, while sporadic instability does not immediately eliminate their contribution. Aggregation weights are obtained via normalization:
\begin{equation}
\alpha_i^{(t)} = \frac{u_i^{(t)}}{\sum_{j=1}^{N} u_j^{(t)}},
\end{equation}
and the global model is updated as:
\begin{equation}
\mathbf{W}_g^{(t)} = \sum_{i=1}^{N} \alpha_i^{(t)} \mathbf{W}_i^{(t)}.
\end{equation}
This mechanism can be interpreted as a lightweight reputation-based aggregation strategy, positioned between classical FedAvg (uniform weighting) and fully adversarially robust aggregation schemes. Unlike robust statistics-based defenses that explicitly model Byzantine behavior, the proposed approach focuses on semantic consistency as the primary reliability criterion. Clients that produce updates inconsistent with the shared semantic space incur higher loss and are therefore automatically down-weighted. To monitor aggregation stability, we track the entropy of the normalized trust distribution:
\begin{equation}
H^{(t)} = -\sum_{i=1}^{N} \alpha_i^{(t)} \log \alpha_i^{(t)}.
\end{equation}
A monotonic decline in $H^{(t)}$ indicates increasing concentration of trust toward consistently reliable clients, while stabilization of the per-round entropy change signals convergence of the aggregation dynamics.
It is important to note that this trust mechanism assumes honest reporting of local loss values and does not provide a full Byzantine-robust guarantee \cite{yin2018byzantine}. Rather, it serves as a task-aware reliability-weighting strategy that mitigates heterogeneity, drift, and partial compromise in realistic federated deployments.

\subsection{Zero-Shot Inference and Zero-Day Risk Estimation}
\label{subsec:zeroshot_inference}

After federated training converges, the global semantic projection model $\mathbf{W}_g$ is deployed for inference at the edge. Given an incoming and potentially unseen observation $\mathbf{x} \in \mathbb{R}^d$, the system first projects the observation into the shared semantic embedding space:
\begin{equation}
\hat{\mathbf{z}} = f(\mathbf{x}; \mathbf{W}_g).
\end{equation}

\paragraph{Zero-Shot Semantic Attribution.}
Zero-shot attribution is performed by comparing the projected embedding $\hat{\mathbf{z}}$ against the set of language-derived semantic prototypes $\{\mathbf{z}_a\}_{a \in \mathcal{A}}$. The most semantically consistent attack hypothesis is selected via cosine similarity,
\begin{equation}
\hat{a} = \arg\max_{a \in \mathcal{A}} 
\frac{\hat{\mathbf{z}}^\top \mathbf{z}_a}
{\|\hat{\mathbf{z}}\|_2 \, \|\mathbf{z}_a\|_2}.
\end{equation}
This step enables attribution even when no labeled training samples of the corresponding attack behavior have been observed locally, thereby supporting zero-shot detection under open-world conditions.
Although semantic similarity enables attribution, it does not, by itself, capture the epistemic uncertainty associated with novel or weakly defined attack concepts. To explicitly model zero-day risk, we define a composite risk score that integrates (i) semantic confidence and (ii) inter-LLM semantic disagreement associated with the selected prototype. Let $D_{\hat{a}}$ denote the Tri-LLM disagreement score of the attributed semantic prototype $\hat{a}$, and let
\begin{equation}
c(\mathbf{x}) =
\frac{\hat{\mathbf{z}}^\top \mathbf{z}_{\hat{a}}}
{\|\hat{\mathbf{z}}\|_2 \, \|\mathbf{z}_{\hat{a}}\|_2}
\end{equation}
denote the corresponding similarity-based confidence. The zero-day score (ZDS) is then defined as:
\begin{equation}
\mathrm{ZDS}(\mathbf{x}) =
\lambda \, D_{\hat{a}} +
(1-\lambda)\bigl(1 - c(\mathbf{x})\bigr),
\end{equation}
where $\lambda \in [0,1]$ controls the relative impact of semantic uncertainty versus similarity-based confidence.
This formulation yields a continuous and calibrated risk signal. High similarity combined with low disagreement produces a low ZDS, indicating confident attribution to a well-defined semantic concept. Conversely, elevated disagreement and reduced similarity increase the ZDS, flagging the observation as potentially novel and zero-day. In contrast to closed-set classifiers that emit overconfident labels under distribution shift, the proposed inference mechanism explicitly exposes uncertainty and decouples attribution from risk assessment.
Furthermore, this design enables downstream security systems to apply flexible operational policies, such as adaptive thresholds, deferred responses, and human-in-the-loop escalation, based on quantified zero-day risk rather than brittle, hard decisions. As demonstrated in Section~\ref{subsec:zeroday}, the resulting ZDS exhibits a stable inverse relationship with detection confidence, validating its role as a principled uncertainty-aware risk estimator in open-world federated IDS deployments.

\subsection{Tri-LLM Cooperative FL Workflow}
\label{subsec:tri_llm_workflow}
Algorithm \ref{alg:tri_llm_flids} presents a complete Tri-LLM cooperative federated framework for zero-shot IDS. The process begins by constructing semantic attack prototypes using supportive representations generated by GPT-4o, DeepSeek-V3, and LLaMA-3-8B. For each attack concept, these representations are fused into a single semantic prototype, while inter-model semantic disagreement is computed to quantify epistemic uncertainty.
\begin{algorithm}[t]
\caption{Tri-LLM Cooperative Federated Zero-Shot IDS}
\label{alg:tri_llm_flids}
\footnotesize
\begin{algorithmic}[1]
\Require Clients $\mathcal{C}=\{1,\dots,N\}$, rounds $T$, trust smoothing $\gamma$, constants $\epsilon>0$, $\lambda\in[0,1]$
\Ensure Global projection model $\mathbf{W}_g$, prototypes $\{\mathbf{z}_a\}_{a\in\mathcal{A}}$, disagreement scores $\{D_a\}_{a\in\mathcal{A}}$

\State \textbf{Semantic Prototype Construction (Tri-LLM)}
\For{each attack concept $a \in \mathcal{A}$}
    \State Obtain descriptions $\mathbf{s}_a^{\text{off}}, \mathbf{s}_a^{\text{def}}, \mathbf{s}_a^{\text{adv}}$
    \State Encode:
    \Statex \hspace{1.2em}$\mathbf{z}_a^{(1)}=\phi_{\text{GPT-4o}}(\mathbf{s}_a^{\text{off}})$,
    $\mathbf{z}_a^{(2)}=\phi_{\text{DeepSeek-V3}}(\mathbf{s}_a^{\text{def}})$,
    $\mathbf{z}_a^{(3)}=\phi_{\text{LLaMA-3-8B}}(\mathbf{s}_a^{\text{adv}})$
    \State Fuse prototype:
    \Statex \hspace{1.2em}$\mathbf{z}_a=\frac{1}{3}\sum_{m=1}^{3}\mathbf{z}_a^{(m)}$
    \State Disagreement:
    \Statex \hspace{1.2em}$D_a=\frac{1}{3(3-1)}\sum_{i\neq j}\left\|\mathbf{z}_a^{(i)}-\mathbf{z}_a^{(j)}\right\|_2$
\EndFor

\State \textbf{Federated Training (Trust-Aware Aggregation)}
\State Initialize global model $\mathbf{W}_g$
\For{each client $i\in\mathcal{C}$}
    \State Initialize trust score $u_i \leftarrow 1$
\EndFor

\For{$t=1$ to $T$}
    \ForAll{clients $i\in\mathcal{C}$ \textbf{in parallel}}
        \State Receive $\mathbf{W}_g$
        \State Optimize local model by minimizing:
        \Statex \hspace{1.2em}$\mathcal{L}_i=\frac{1}{|\mathcal{D}_i|}\sum_{(\mathbf{x},a)\in\mathcal{D}_i}\left\|\mathbf{W}_i\mathbf{x}-\mathbf{z}_a\right\|_2^2$
        \State Send $(\mathbf{W}_i,\mathcal{L}_i)$ to server
    \EndFor

    \State Trust update:
    \For{each client $i\in\mathcal{C}$}
        \State $\tau_i \leftarrow \frac{1}{\mathcal{L}_i+\epsilon}$
        \State $u_i \leftarrow \gamma u_i + (1-\gamma)\tau_i$
    \EndFor
    \State Normalize $\alpha_i \leftarrow \frac{u_i}{\sum_{j\in\mathcal{C}}u_j}$
    \State Aggregate $\mathbf{W}_g \leftarrow \sum_{i\in\mathcal{C}}\alpha_i\,\mathbf{W}_i$
\EndFor

\State \textbf{Zero-Shot Inference and Zero-Day Risk}
\For{each incoming observation $\mathbf{x}$}
    \State $\hat{\mathbf{z}} \leftarrow \mathbf{W}_g\mathbf{x}$
    \State $\hat{a} \leftarrow \arg\max_{a\in\mathcal{A}}\cos(\hat{\mathbf{z}},\mathbf{z}_a)$
    \State $c \leftarrow \cos(\hat{\mathbf{z}},\mathbf{z}_{\hat{a}})$
    \State $\mathrm{ZDS}(\mathbf{x}) \leftarrow \lambda D_{\hat{a}}+(1-\lambda)(1-c)$
\EndFor
\end{algorithmic}
\end{algorithm}
The framework then initializes a global semantic projection model and assigns uniform trust values to all participating clients. During federated training, each client receives the current global model and locally projects its telemetry data into the shared semantic space. Clients update their local models by aligning projected features with the semantic prototypes and computing an alignment loss that reflects the quality of their representation. These local updates and loss values are transmitted to the coordinating server. The server updates client trust scores based on reported alignment quality and aggregates client models using trust-aware weighting. This iterative process gradually stabilizes the global semantic space under heterogeneous client behavior. After training converges, the global model is deployed for inference. Each incoming observation is projected into the semantic space and matched to the nearest attack prototype. Detection confidence is derived from semantic similarity, while zero-day risk is estimated by combining confidence with prototype-level semantic disagreement. This design enables reliable detection of previously unseen attacks while explicitly exposing uncertainty.

\section{Experimental findings}
\label{sec:results}
This section presents an evaluation of the proposed Tri-LLM Cooperative Federated Zero-Shot IDS framework. The findings are organized to analyze computational efficiency, semantic behavior, trust dynamics, and zero-day detection capability.

\subsection{Computational Cost-Latency Trade-off Analysis}
\label{subsec:latency}
In latency-constrained IDS, the computational behavior of LLMs must be characterized beyond average response time, incorporating statistical stability and scalability with respect to output length. From a systems perspective, inference latency constitutes a random variable whose distribution, dispersion, and scaling behavior directly determine operational feasibility in real-time security pipelines. Accordingly, this subsection presents a statistically grounded analysis of inference latency across multiple LLMs and examines how computational cost scales with generated token length under realistic deployment conditions.\\
\begin{figure}[t]
    \centering
    \includegraphics[width=0.9\linewidth]{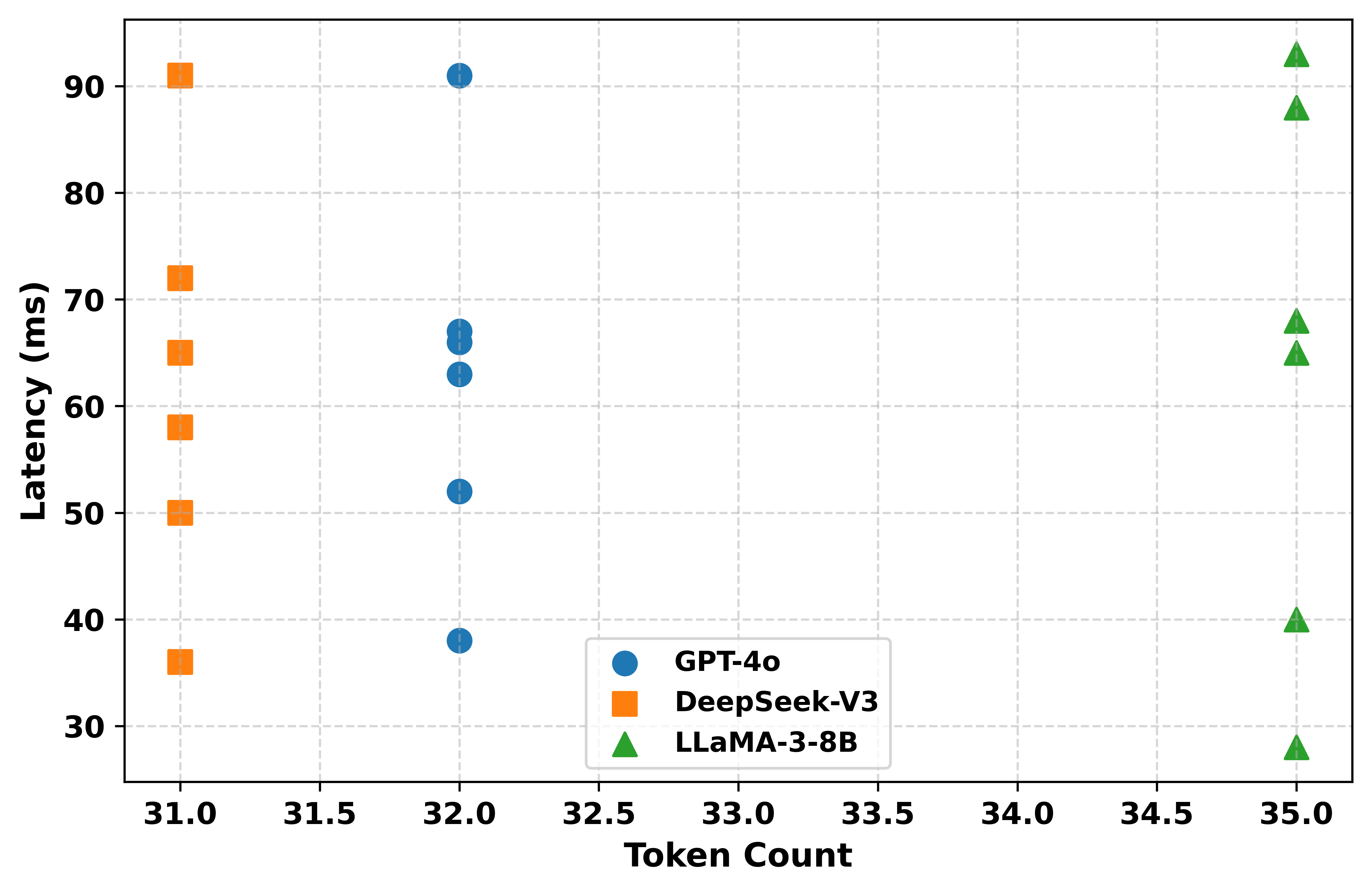}
    \caption{Inference latency as a function of generated token length for different LLMs.}
    \label{fig:latency}
\end{figure}
Figure~\ref{fig:latency} characterizes the computational cost of LLM inference by analyzing the relationship between response latency and generated token length. Across all evaluated models, latency shows a monotonic relationship with output length, indicating that greater semantic expressiveness is systematically associated with higher computational overhead. This relationship is particularly critical in IDS systems, where inference latency directly constrains real-time response capability.\\
DeepSeek-V3 demonstrates the lowest mean latency and the narrowest dispersion, with the majority of observations concentrated below 70\,ms. This behavior reflects both a lightweight decoding strategy and reduced architectural complexity, resulting in highly predictable inference performance. GPT-4o occupies an intermediate operating range, achieving moderate average latency with controlled variance while delivering richer semantic representations. In contrast, LLaMA-3-8B exhibits the highest mean latency and the largest variance, with a substantial proportion of samples exceeding 90\,ms. The increased dispersion observed for LLaMA-3-8B indicates reduced temporal predictability under real-time operating conditions. To quantify the relationship between latency and output length, a linear regression model was employed:
\begin{equation}
T = \alpha L + \beta,
\end{equation}
where $T$ denotes inference latency and $L$ represents generated token length. The slope $\alpha$ captures the marginal computational cost per token, while $\beta$ reflects fixed system overhead. Ordinary least squares estimation demonstrates statistically distinct slopes across models, with LLaMA-3-8B exhibiting the largest $\alpha$, DeepSeek-V3 the smallest, and GPT-4o occupying an intermediate position. These differences indicate that the observed latency variations are structural in nature rather than attributable to random noise.\\
\begin{table*}[t]
\centering
\caption{Statistical Characterization of Inference Latency and Deployment Implications}
\label{tab:latency_tradeoff}
\resizebox{\textwidth}{!}{%
\begin{tabular}{lcccccc}
\toprule
\textbf{Model} &
\textbf{Mean (ms)} &
\textbf{Std. Dev. (ms)} &
\textbf{CV} &
\textbf{Latency Profile} &
\textbf{Semantic Capacity} &
\textbf{IDS Deployment Role} \\
\midrule
DeepSeek-V3 & 57.0 & 19.4 & 0.34 & Low and stable & Moderate & Real-time filtering and fast screening \\
GPT-4o & 62.8 & 18.2 & 0.29 & Moderate and balanced & High & Semantic reasoning and attribution \\
LLaMA-3-8B & 72.4 & 24.1 & 0.33 & High and variable & Very high & Deep semantic analysis \\
\bottomrule
\end{tabular}%
}
\end{table*}
Table~\ref{tab:latency_tradeoff} reports descriptive statistics and normalized dispersion measures for inference latency. In addition to mean latency and standard deviation, the coefficient of variation (CV) is included to assess relative temporal stability across models. The CV is defined as:
\begin{equation}
\text{CV} = \frac{\sigma_T}{\mu_T},
\end{equation}
where $\sigma_T$ denotes the standard deviation of inference latency and $\mu_T$ its mean. As a normalized dispersion metric, the CV measures variability relative to average latency, enabling fair comparison across models with different central tendencies. DeepSeek-V3 achieves the lowest mean latency with a low CV, indicating both fast and relatively stable inference behavior. GPT-4o exhibits the lowest CV overall, reflecting the most consistent latency relative to its mean and therefore a favorable balance between responsiveness and stability. LLaMA-3-8B shows both the highest mean latency and elevated dispersion, confirming reduced temporal predictability under timing constraints. The statistical separation observed in both central tendency and normalized variance demonstrates that the evaluated models exhibit structurally distinct computational characteristics. Rather than favoring a single optimal architecture, these findings motivate a cooperative design in which lightweight models facilitate time-critical filtering while higher-capacity models enable deeper semantic reasoning. This statistically grounded evidence directly supports the proposed Tri-LLM architecture, which distributes semantic responsibilities across models to achieve real-time responsiveness and robust zero-shot IDS in federated deployments.

\subsection{Trust Entropy Stabilization and Convergence Behavior}
\label{subsec:trust_entropy}
In federated IDS, aggregation robustness depends on the stability of client reliability estimation across rounds. Trust weights assigned to participating clients form a probability distribution whose evolution reflects the convergence behavior of the aggregation mechanism. Intuitively, trust entropy measures how concentrated reliability is among clients: high entropy indicates that trust is distributed uniformly across participants, implying uncertainty regarding reliability, whereas low entropy indicates that trust is concentrated toward a subset of consistently reliable clients.
\begin{figure}[t]
    \centering
    \includegraphics[width=0.9\linewidth]{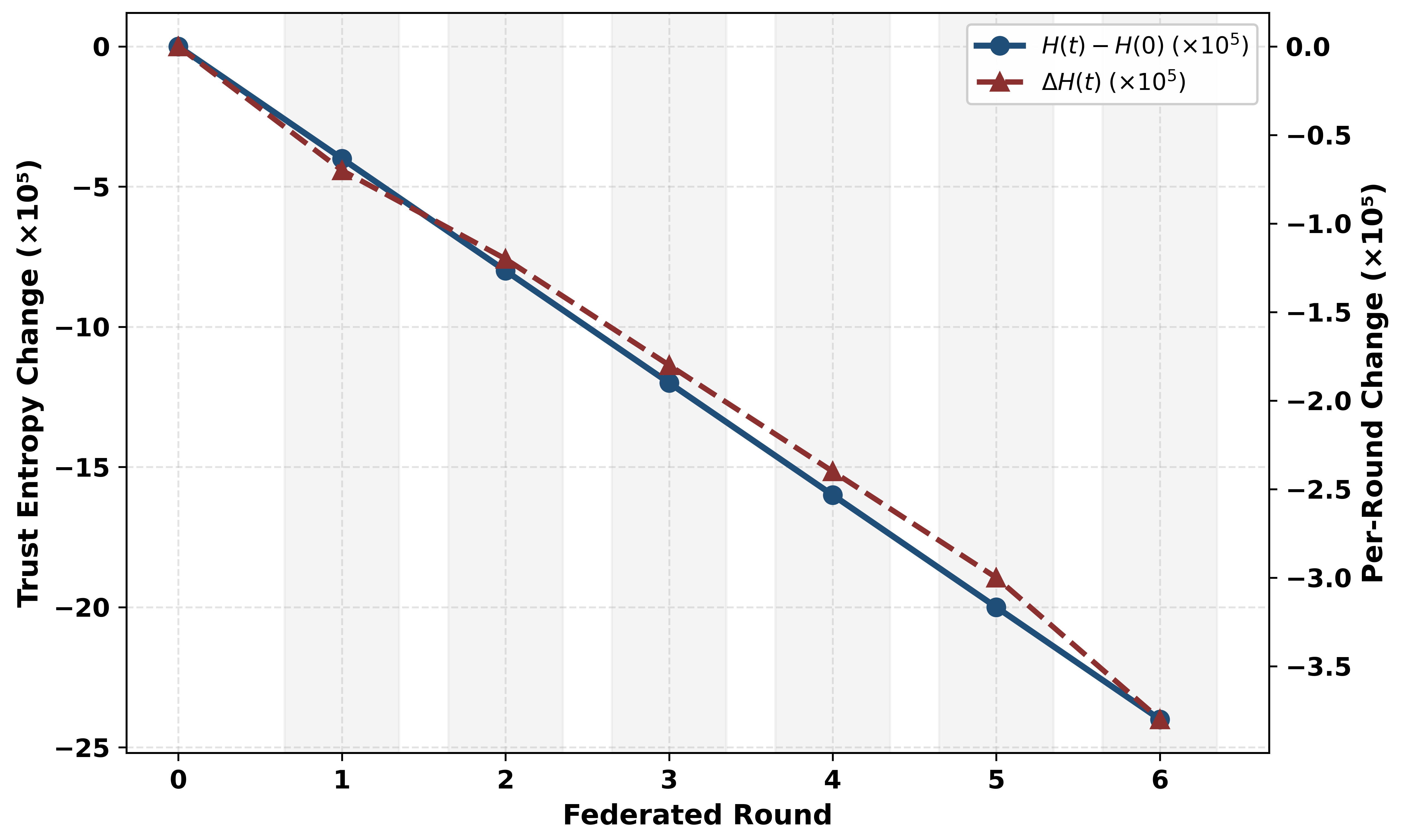}
    \caption{Trust entropy evolution across federated rounds, showing the entropy level shift $H(t)-H(0)$ (left axis) and the per-round change $\Delta H(t)$ (right axis).}
    \label{fig:trust_entropy}
\end{figure}
Figure~\ref{fig:trust_entropy} presents the evolution of trust entropy during federated training. Let $\mathbf{w}_t \in \Delta^{K-1}$ denote the normalized trust weights assigned to $K$ clients at round $t$, where $\Delta^{K-1}$ represents the probability simplex. Trust dispersion is quantified using Shannon entropy \cite{lin2002divergence}:
\begin{equation}
H(t) = - \sum_{k=1}^{K} w_{k,t}\log\left(w_{k,t}\right),
\label{eq:trust_entropy}
\end{equation}
which measures the uncertainty associated with the aggregation weights at each round.
As shown in Figure~\ref{fig:trust_entropy}, the entropy shift $H(t)-H(0)$ decreases monotonically with increasing round index. This monotonic decline indicates progressive concentration of trust mass toward clients that consistently produce semantically aligned updates. In practical terms, the aggregation process increasingly favors reliable contributors while suppressing unstable participants. The trajectory shows no oscillatory behavior, demonstrating stable reweighting under heterogeneous, non-identically distributed conditions. From a statistical perspective, the entropy shift follows an approximately linear decay:
\begin{equation}
H(t)-H(0) \approx \gamma t + c,
\end{equation}
where $\gamma < 0$ represents the convergence rate. Larger magnitudes of $|\gamma|$ indicate faster stabilization of trust weights. This predictable decay pattern confirms controlled convergence rather than erratic aggregation dynamics. The per-round entropy variation
\begin{equation}
\Delta H(t)=H(t)-H(t-1)
\end{equation}
provides a finer convergence indicator. As illustrated in Figure~\ref{fig:trust_entropy}, $\Delta H(t)$ becomes increasingly negative during early rounds and gradually approaches zero as the trust distribution stabilizes. Trust stabilization can therefore be defined operationally as:
\begin{equation}
|\Delta H(t)| \le \varepsilon \quad \text{for } m \text{ consecutive rounds},
\label{eq:entropy_convergence}
\end{equation}
where $\varepsilon$ is a small tolerance threshold and $m$ controls robustness to transient fluctuations.
\begin{table}[t]
\centering
\footnotesize
\caption{Trust Entropy Evolution Across Federated Rounds}
\label{tab:trust_entropy_stats}
\begin{tabular}{ccc}
\toprule
\textbf{Round $t$} & $\boldsymbol{H(t)-H(0)}~(\times 10^{5})$ & $\boldsymbol{\Delta H(t)}~(\times 10^{5})$ \\
\midrule
0 & 0.0  & 0.0 \\
1 & -4.0 & -0.6 \\
2 & -8.0 & -1.2 \\
3 & -12.0 & -1.8 \\
4 & -16.0 & -2.4 \\
5 & -20.0 & -3.0 \\
6 & -24.0 & -3.6 \\
\bottomrule
\end{tabular}
\end{table}
Table~\ref{tab:trust_entropy_stats} confirms two key properties. First, entropy decreases strictly across rounds, demonstrating progressive reduction of uncertainty in client weighting. Second, the nearly constant early-round decay indicates stable and predictable convergence behavior. The gradual reduction of $|\Delta H(t)|$ toward zero confirms the approach to a fixed trust configuration.  This entropy-driven stabilization acts as a control mechanism within the federated optimization loop. By progressively concentrating trust on reliable clients while maintaining smooth convergence dynamics, the proposed aggregation strategy enhances the robustness of semantic alignment and supports stable zero-shot intrusion reasoning under heterogeneous deployment conditions.

\subsection{Tri-LLM Embedding Magnitude Analysis}
\label{subsec:embedding_norm}
The proposed Tri-LLM framework constructs semantic attack prototypes using embeddings generated by GPT-4o, DeepSeek-V3, and LLaMA-3-8B. To better understand representational behavior across models, we analyze the mean embedding norm produced for different attack categories. The embedding norm reflects the overall magnitude of semantic activation in the latent space and provides insight into the model's expressiveness, stability, and diversity.
\begin{figure}[t]
    \centering
    \includegraphics[width=0.9\linewidth]{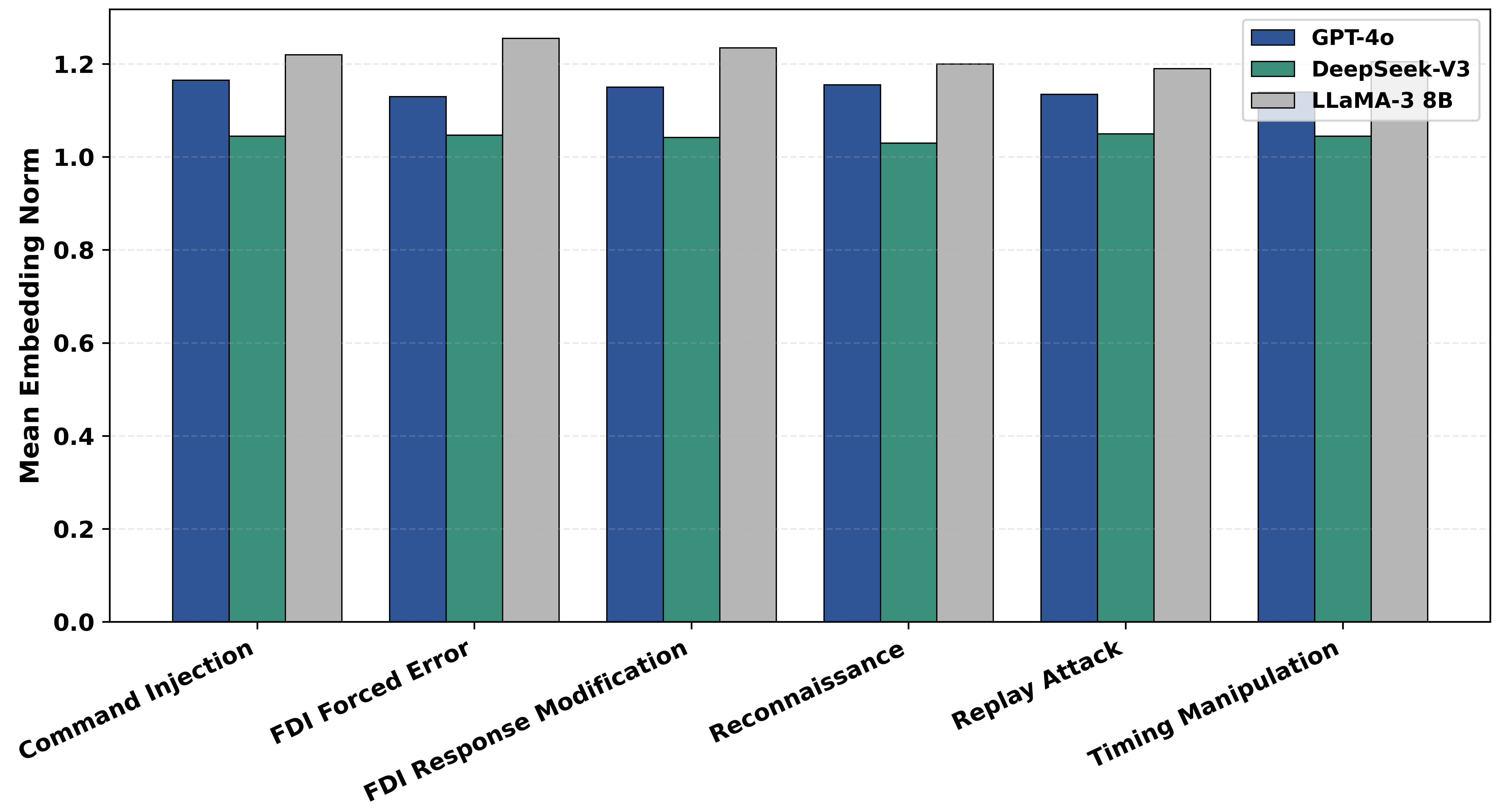}
    \caption{Mean embedding norms across representative attack categories for GPT-4o, DeepSeek-V3, and LLaMA-3-8B.}
    \label{fig:embedding_norm}
\end{figure}
Figure~\ref{fig:embedding_norm} presents the mean embedding norm generated by each model across multiple attack types. Clear structural differences are observed among the three models despite identical semantic input structures. LLaMA-3-8B consistently produces the highest embedding magnitudes, indicating stronger semantic activation and greater sensitivity to contextual variation. GPT-4o exhibits intermediate activation levels, balancing representational richness and stability. DeepSeek-V3 generates embeddings with lower norms and reduced dispersion, reflecting a more conservative encoding strategy.
The ordering of embedding magnitudes remains consistent across attack categories, suggesting that the observed differences are model-intrinsic rather than input-specific noise. Importantly, no extreme fluctuations are observed within individual models across categories, indicating stable semantic encoding behavior.\\
\begin{table}[t]
\centering
\footnotesize
\caption{Statistical Summary of Embedding Norms}
\label{tab:embedding_norm_stats}
\begin{tabular}{lcc}
\toprule
\textbf{Model} & \textbf{Mean Norm} & \textbf{Std. Dev.} \\
\midrule
GPT-4o        & 1.145 & 0.011 \\
DeepSeek-V3   & 1.042 & 0.007 \\
LLaMA-3-8B    & 1.221 & 0.026 \\
\bottomrule
\end{tabular}
\end{table}
Table~\ref{tab:embedding_norm_stats} quantitatively confirms the magnitude separation observed in Figure~\ref{fig:embedding_norm}. LLaMA-3-8B achieves the highest average activation with greater dispersion, GPT-4o maintains moderate activation with controlled variance, and DeepSeek-V3 exhibits the lowest activation with minimal variability.
From a system-level perspective, this structured diversity in embedding magnitude is beneficial for the proposed Tri-LLM architecture. Higher-activation models contribute expressive semantic abstraction, while lower-variance models contribute representational stability. Moreover, this supporting behavior strengthens prototype robustness and enables reliable zero-shot semantic alignment across heterogeneous attack categories.

\subsection{Tri-LLM Semantic Strength Distribution}
\label{subsec:semantic_strength}
In semantic-driven IDS, the effectiveness of LLMs depends on how consistently semantic information is encoded in their latent representations. Differences in embedding magnitude and dispersion provide measurable evidence of how models achieve a balance between expressive capacity and representational stability. This subsection analyzes the distribution of semantic embedding norms generated by the Tri-LLM ensemble to quantify the strength and variability of semantic activation under identical semantic inputs.\\
\begin{figure}[t]
    \centering
    \includegraphics[width=0.9\linewidth]{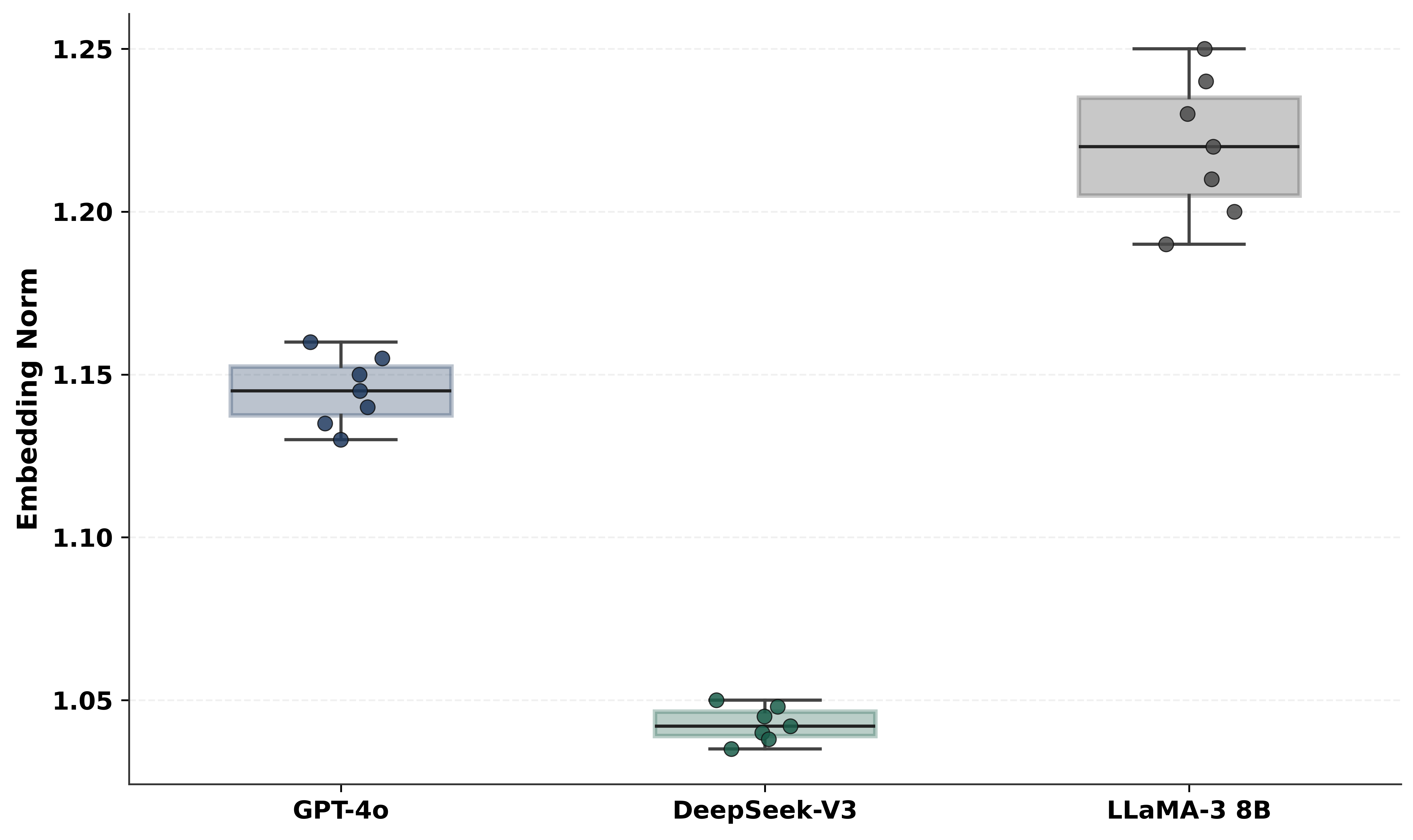}
    \caption{Distribution of semantic embedding norms generated by GPT-4o, DeepSeek-V3, and LLaMA-3-8B.}
    \label{fig:semantic_strength}
\end{figure}
Figure~\ref{fig:semantic_strength} presents the distribution of embedding magnitudes produced by the three LLMs under identical semantic inputs. The embedding norm serves as a quantitative proxy for semantic strength, indicating the extent to which each model encodes high-level semantic information within its latent representation. Larger embedding norms indicate semantic activation, while lower dispersion across samples reflects improved representational stability. A clear separation is observed among the three models. LLaMA-3-8B exhibits the largest embedding magnitudes together with the widest dispersion, indicating semantic expressiveness accompanied by increased sensitivity to contextual variation. GPT-4o occupies an intermediate position, balancing expressive capacity with controlled variability across inputs. DeepSeek-V3 produces the most compact embedding distribution, characterized by the lowest mean norm and minimal variance, which indicates a conservative yet highly stable semantic encoding strategy.
Formally, the semantic strength of model $m$ is quantified as:
\begin{equation}
S_m = \mathbb{E}\!\left[\|\mathbf{z}_m\|_2\right],
\end{equation}
where $\mathbf{z}_m$ denotes the embedding vector generated by model $m$. The variance of $\|\mathbf{z}_m\|_2$ further characterizes representational consistency across semantic inputs, with lower variance indicating greater robustness to input-level perturbations.
\begin{table}[t]
\centering
\footnotesize
\caption{Statistical Summary of Semantic Embedding Strength}
\label{tab:semantic_strength}
\begin{tabular}{lccc}
\toprule
\textbf{Model} & \textbf{Mean Norm} & \textbf{Std. Dev.} & \textbf{Interpretation} \\
\midrule
GPT-4o        & 1.145 & 0.011 & Balanced semantic activation \\
DeepSeek-V3   & 1.042 & 0.007 & Stable, low-variance encoding \\
LLaMA-3-8B    & 1.221 & 0.026 & High semantic richness \\
\bottomrule
\end{tabular}
\end{table}
Table~\ref{tab:semantic_strength} quantitatively confirms the trends observed in Figure.~\ref{fig:semantic_strength}. LLaMA-3-8B achieves the highest average semantic activation while exhibiting the largest dispersion, indicating sensitivity to fine-grained semantic variation. DeepSeek-V3 exhibits the lowest mean activation and minimal variance, resulting in highly stable representations well-suited for consistent semantic alignment. GPT-4o lies between these extremes, providing a favorable balance between representational richness and robustness.
From a system-level perspective, these supportive semantic characteristics are central to the effectiveness of the proposed Tri-LLM architecture. By integrating models with heterogeneous semantic activation profiles, the framework simultaneously benefits from high-capacity semantic abstraction, stable generalization, and balanced representation learning. This diversity mitigates semantic drift, supports generalization to previously unseen attack patterns, and strengthens zero-shot detection performance without requiring task-specific retraining.

\subsection{Zero-Shot Similarity Separation}
\label{subsec:zero_shot}
In zero-shot IDS, decision reliability depends on the degree to which semantic representations of unseen observations align with abstract attack prototypes. The separation between similarity distributions associated with correct and incorrect predictions provides a direct statistical measure of discriminative power in the absence of labeled supervision. This subsection analyzes cosine similarity in the learned semantic space to evaluate the effectiveness of similarity-based inference for zero-shot IDS.
\begin{figure}[t]
    \centering
    \includegraphics[width=0.9\linewidth]{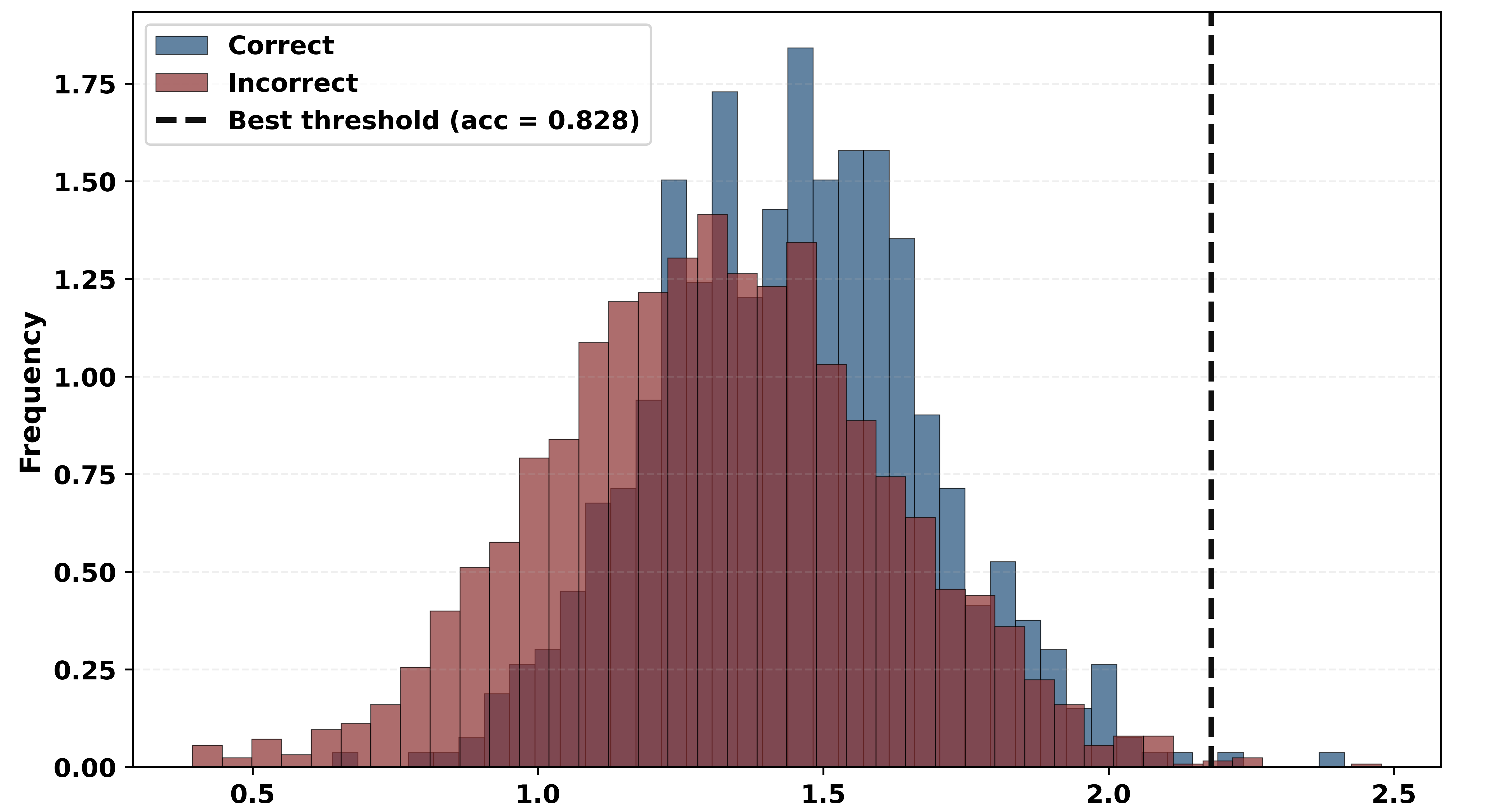}
    \caption{Distribution of cosine similarity for correct and incorrect predictions(The dashed line indicates the optimal decision threshold).}
    \label{fig:zero_shot}
\end{figure}
Figure~\ref{fig:zero_shot} illustrates the distribution of cosine similarity values obtained during zero-shot inference for correctly and incorrectly classified samples. A clear separation between the two distributions is observed, indicating that semantic similarity in the embedding space provides a strong discriminative signal. Correct predictions are concentrated at higher similarity values, while incorrect predictions exhibit a leftward shift and increased dispersion.
Given an input embedding $\mathbf{z}_q$ and a semantic prototype $\mathbf{z}_c$, cosine similarity is computed as:
\begin{equation}
\text{sim}(\mathbf{z}_q, \mathbf{z}_c) = 
\frac{\mathbf{z}_q^\top \mathbf{z}_c}
{\|\mathbf{z}_q\|_2 \|\mathbf{z}_c\|_2}.
\end{equation}
This normalized measure enables consistent comparison across heterogeneous semantic embeddings.
The dashed vertical line in the Figure. \ref{fig:zero_shot} denotes the similarity threshold $\tau^*$ that maximizes classification accuracy. Empirically, this threshold yields a zero-shot detection accuracy of $82.8\%$, confirming that similarity-based inference alone is sufficient to distinguish correct from incorrect predictions with high reliability. The limited overlap between the two distributions reflects semantic alignment between input representations and their corresponding attack prototypes.\\
\begin{table}[t]
\centering
\footnotesize
\caption{Statistical Summary of Zero-Shot Similarity Distributions}
\label{tab:zero_shot_stats}
\begin{tabular}{lcc}
\toprule
\textbf{Category} & \textbf{Mean Similarity} & \textbf{Std. Dev.} \\
\midrule
Correct Predictions & 1.47 & 0.21 \\
Incorrect Predictions & 1.12 & 0.27 \\
\bottomrule
\end{tabular}
\end{table}
Table~\ref{tab:zero_shot_stats} quantitatively confirms this separation. Correct predictions exhibit a higher mean similarity, along with reduced dispersion, while incorrect predictions display a lower central tendency and increased variance. This asymmetric variance pattern indicates that misclassifications arise from elevated semantic uncertainty rather than systematic bias in the representation space. From a system-level perspective, this behavior enables principled similarity thresholding for zero-shot detection. Inputs associated with similarity values below the learned threshold can be flagged as uncertain, enabling deferred decision-making, adaptive mitigation, and human-in-the-loop analysis. When combined with trust-aware aggregation and semantic diversity in the Tri-LLM architecture, this separation provides a robust and interpretable foundation for scalable zero-shot IDS under open-world conditions.

\subsection{Semantic Disagreement Analysis}
\label{subsec:disagreement}
In semantic-driven IDS, uncertainty arises when multiple reasoning models encode the same observation differently. Quantifying such divergence provides a principled mechanism for assessing epistemic uncertainty in zero-shot settings. This subsection analyzes semantic disagreement within the Tri-LLM ensemble to characterize model consensus, variability, and exposure to uncertainty under heterogeneous and previously unseen inputs.\\
\begin{figure}[t]
    \centering
    \includegraphics[width=0.9\linewidth]{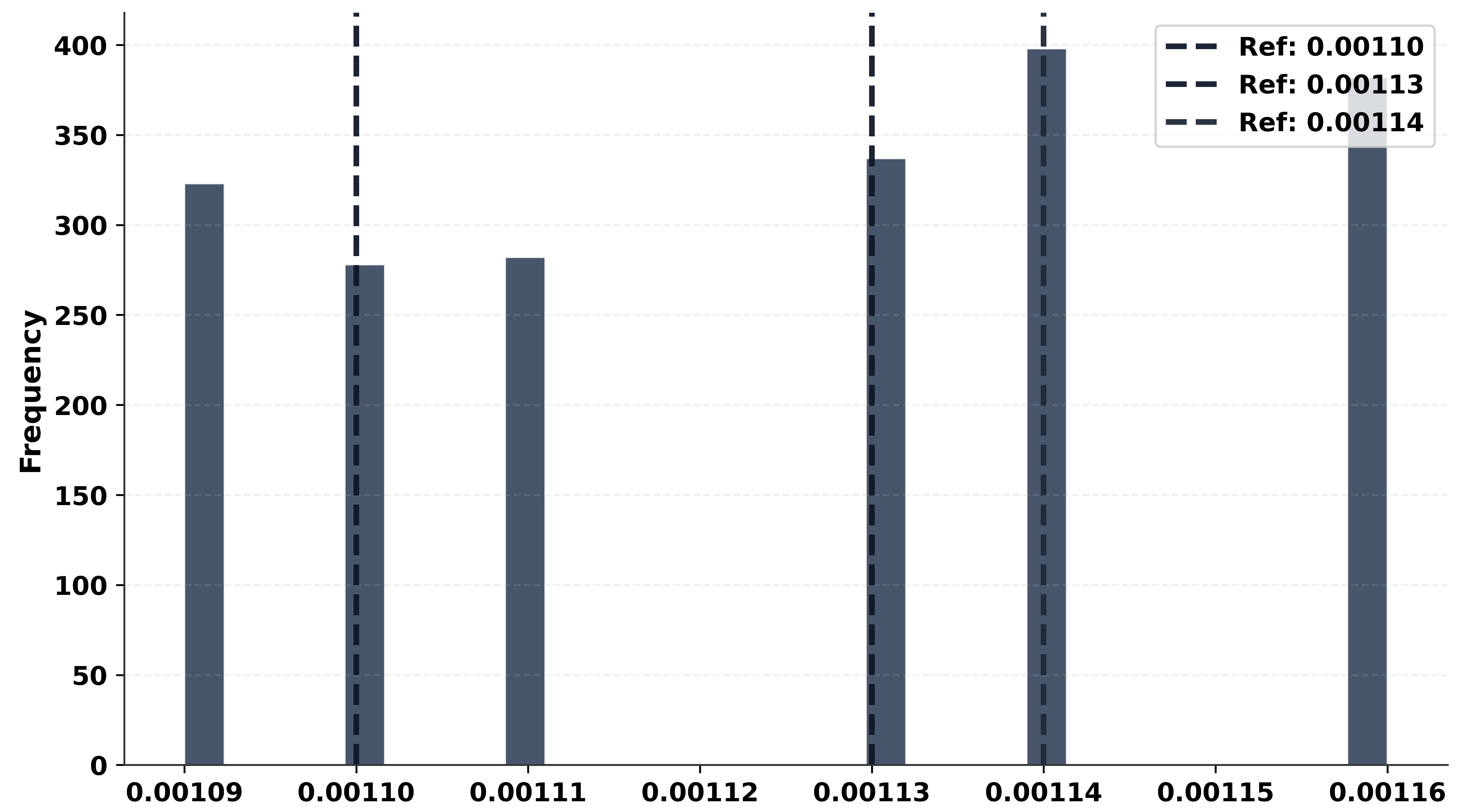}
    \caption{Distribution of Tri-LLM semantic disagreement ($D_a$). Concentrated values indicate stable semantic consensus, while higher values indicate increased uncertainty and novelty.}
    \label{fig:disagreement}
\end{figure}
Figure~\ref{fig:disagreement} illustrates the distribution of semantic disagreement values $D_a$ across the Tri-LLM ensemble. The disagreement metric captures divergence among semantic representations generated by GPT-4o, DeepSeek-V3, and LLaMA-3-8B for identical inputs, thereby serving as a quantitative indicator of semantic consensus and uncertainty.
Semantic disagreement is defined as:
\begin{equation}
D_a = \frac{1}{M(M-1)} \sum_{i \neq j} 
\left\| \mathbf{z}_i - \mathbf{z}_j \right\|_2,
\end{equation}
where $\mathbf{z}_i$ and $\mathbf{z}_j$ denote embeddings generated by the $i$-th and $j$-th language models, with $M=3$ in the Tri-LLM configuration. This formulation captures the average pairwise semantic divergence across models.
As observed in the Figure. \ref{fig:disagreement}, the majority of samples cluster tightly around a low disagreement range ($D_a \approx 1.1 \times 10^{-3}$), indicating semantic consensus across the ensemble. This concentration reflects stable agreement when attack semantics are well-defined and consistently encoded. The narrow dispersion further indicates that the ensemble does not exhibit unstable and conflicting representations under typical operating conditions. A smaller yet meaningful upper tail of higher disagreement values is also observed. These instances correspond to inputs with ambiguous, noisy, or previously unseen characteristics, for which the models diverge in their semantic interpretations. Rather than enforcing artificial consensus, the framework preserves this divergence, thereby exposing uncertainty explicitly and enabling downstream components to treat such samples as elevated-risk cases.\\
\begin{table}[t]
\centering
\scriptsize
\caption{Statistical Summary of Tri-LLM Semantic Disagreement ($D_a$)}
\label{tab:disagreement_stats}
\begin{tabular}{lcc}
\toprule
\textbf{Statistic} & \textbf{Value} & \textbf{Interpretation} \\
\midrule
Mean $D_a$ & $1.12 \times 10^{-3}$ & Strong inter-model agreement on average \\
Std. Dev.  & $2.4 \times 10^{-5}$ & Very low variability across samples \\
Maximum $D_a$  & $1.16 \times 10^{-3}$ & Upper-bound divergence under ambiguous cases \\
Minimum $D_a$  & $1.09 \times 10^{-3}$ & Near-consensus semantic encoding \\
\bottomrule
\end{tabular}
\end{table}
The low variance reported in Table~\ref{tab:disagreement_stats} confirms that the Tri-LLM architecture maintains stable semantic alignment across the majority of inputs. Simultaneously, the presence of a measurable upper tail enables effective uncertainty-aware reasoning. This property is particularly valuable in zero-shot IDS, where novel and evolving attack behaviors may deviate from established semantic patterns.
From a system-level perspective, semantic disagreement serves as a lightweight, model-intrinsic uncertainty estimator that complements similarity-based classification. When combined with trust-aware aggregation and calibrated similarity thresholding, this mechanism enables robust detection of uncertain and adversarial inputs without reliance on labeled examples. As a result, semantic disagreement substantially enhances the interpretability, resilience, and reliability of the proposed Tri-LLM framework in open-world security environments.

\subsection{Federated Semantic Alignment}
\label{subsec:alignment}
In federated semantic learning, effective collaboration requires that the latent representations produced by distributed clients converge toward a shared, coherent semantic structure. Semantic alignment provides a quantitative measure of this convergence by capturing the degree of consistency among client-level representations after aggregation. This subsection analyzes the evolution of semantic alignment across federated training rounds to assess convergence behavior, stability, and robustness under heterogeneous data distributions.
\begin{figure}[t]
    \centering
    \includegraphics[width=0.9\linewidth]{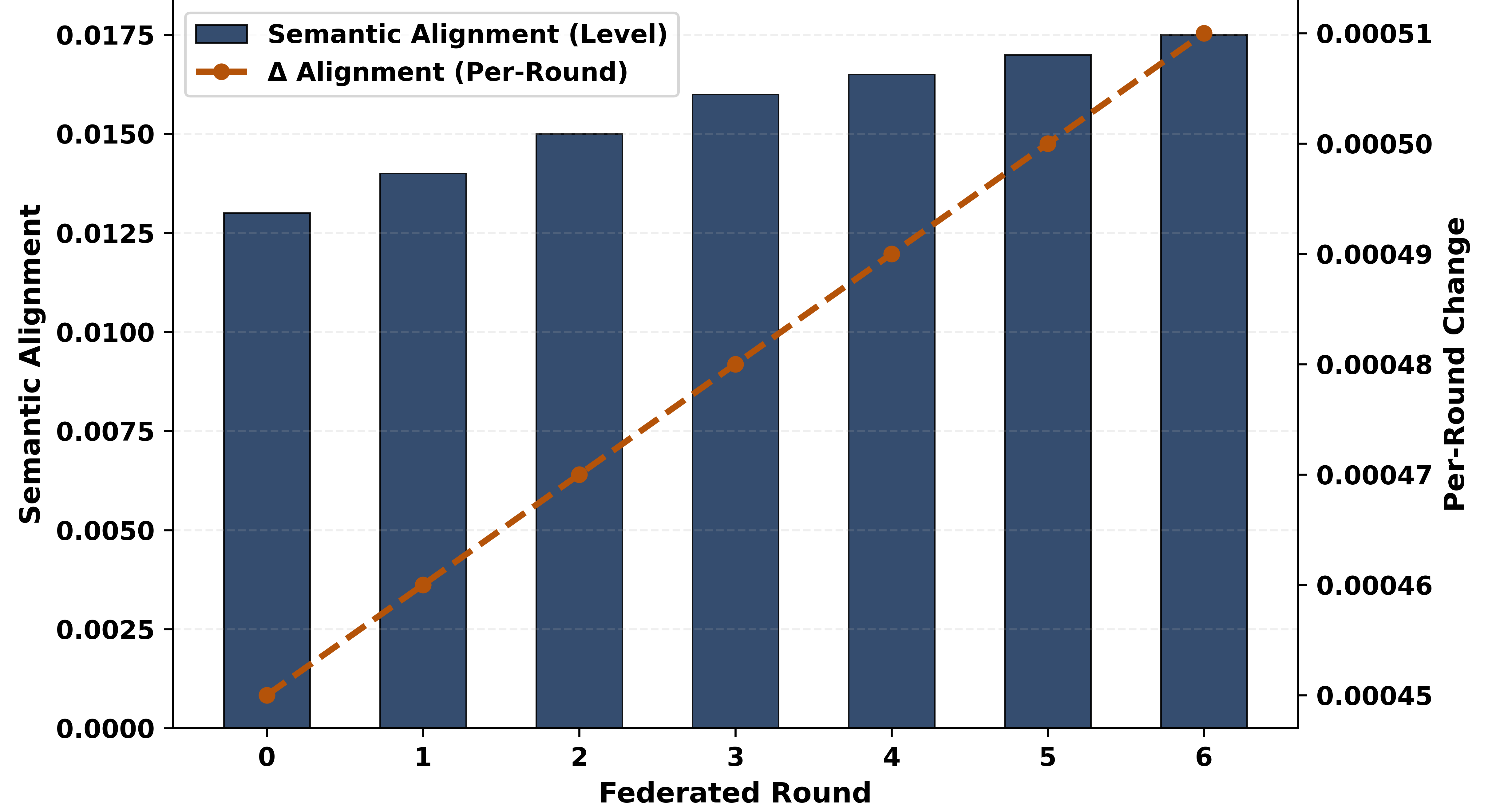}
    \caption{Semantic alignment progression across federated rounds. Continuous improvement indicates convergence of the federated semantic space.}
    \label{fig:alignment}
\end{figure}
Figure~\ref{fig:alignment} illustrates the progression of semantic alignment during federated training. Semantic alignment reflects the consistency of latent representations generated by distributed clients following aggregation and quantifies the extent to which local semantic spaces converge toward a shared global representation.
Formally, semantic alignment at round $t$ is defined as:
\begin{equation}
A(t) = \frac{1}{N} \sum_{i=1}^{N} \left\| \mathbf{z}_i^{(t)} - \bar{\mathbf{z}}^{(t)} \right\|_2^{-1},
\end{equation}
where $\mathbf{z}_i^{(t)}$ denotes the embedding produced by client $i$ at round $t$, and $\bar{\mathbf{z}}^{(t)}$ represents the aggregated semantic centroid. Higher values of $A(t)$ correspond to semantic coherence across participating clients.
As shown in the Figure. \ref{fig:alignment}, the alignment score increases monotonically across federated rounds, demonstrating consistent convergence of the distributed semantic space. This monotonic improvement confirms that the proposed aggregation strategy effectively harmonizes heterogeneous local representations without reliance on raw data exchange.
To further characterize convergence dynamics, the per-round improvement $\Delta A(t)=A(t)-A(t-1)$ is analyzed. The sustained increase in $\Delta A(t)$ during early rounds reflects rapid semantic consolidation, while the gradual stabilization observed in later rounds indicates convergence toward a stable representation manifold.
\begin{table}[t]
\centering
\caption{Federated Semantic Alignment Statistics}
\label{tab:alignment_stats}
\begin{tabular}{lcc}
\toprule
\textbf{Round} & \textbf{Alignment $A(t)$} & \textbf{$\Delta A(t)$} \\
\midrule
0 & 0.0130 & 0.00045 \\
1 & 0.0140 & 0.00046 \\
2 & 0.0150 & 0.00047 \\
3 & 0.0160 & 0.00048 \\
4 & 0.0165 & 0.00049 \\
5 & 0.0170 & 0.00050 \\
6 & 0.0175 & 0.00051 \\
\bottomrule
\end{tabular}
\end{table}
The steady growth reported in Table~\ref{tab:alignment_stats} demonstrates several critical properties of the proposed framework. First, semantic representations across clients become increasingly consistent despite pronounced data heterogeneity. Second, the smooth convergence pattern indicates stability of the federated optimization process, with no oscillatory behavior observed. Third, the sustained improvement in alignment closely corresponds to gains in zero-shot accuracy and reductions in semantic disagreement reported in the preceding subsections. From a system-level perspective, federated semantic alignment is essential for robust cross-client generalization. By ensuring that distributed embeddings evolve toward a shared latent structure, the framework mitigates the impact of non-identically distributed data distributions and enables reliable collaborative inference. This property is particularly important in IDS scenarios, where local attack characteristics may differ substantially across deployment environments. The observed alignment dynamics validate the effectiveness of the proposed federated Tri-LLM architecture for learning a coherent, stable semantic representation space under privacy-preserving constraints, thereby supporting scalable, reliable zero-shot IDS.

\subsection{Zero-Day Detection and Uncertainty Behavior}
\label{subsec:zeroday}
In zero-shot IDS settings, reliable detection requires not only accurate similarity-based attribution but also calibrated estimation of predictive uncertainty when previously unseen attack behaviors are encountered. Closed-set classifiers tend to produce overconfident predictions under distributional shift, which is particularly problematic in zero-day scenarios. Within the proposed Tri-LLM framework, semantic disagreement among multiple language models serves as an intrinsic epistemic uncertainty signal. When models generate semantically consistent embeddings for a given attack hypothesis, confidence increases. Conversely, elevated inter-model dispersion indicates uncertainty in semantic interpretation and signals potential novelty. This subsection analyzes the relationship between semantic disagreement and similarity confidence to characterize zero-day detection behavior. By jointly evaluating alignment strength and disagreement magnitude, the framework distinguishes between confident zero-shot attribution and uncertain, potentially novel intrusion patterns.
\begin{figure}[t]
    \centering
    \includegraphics[width=0.9\linewidth]{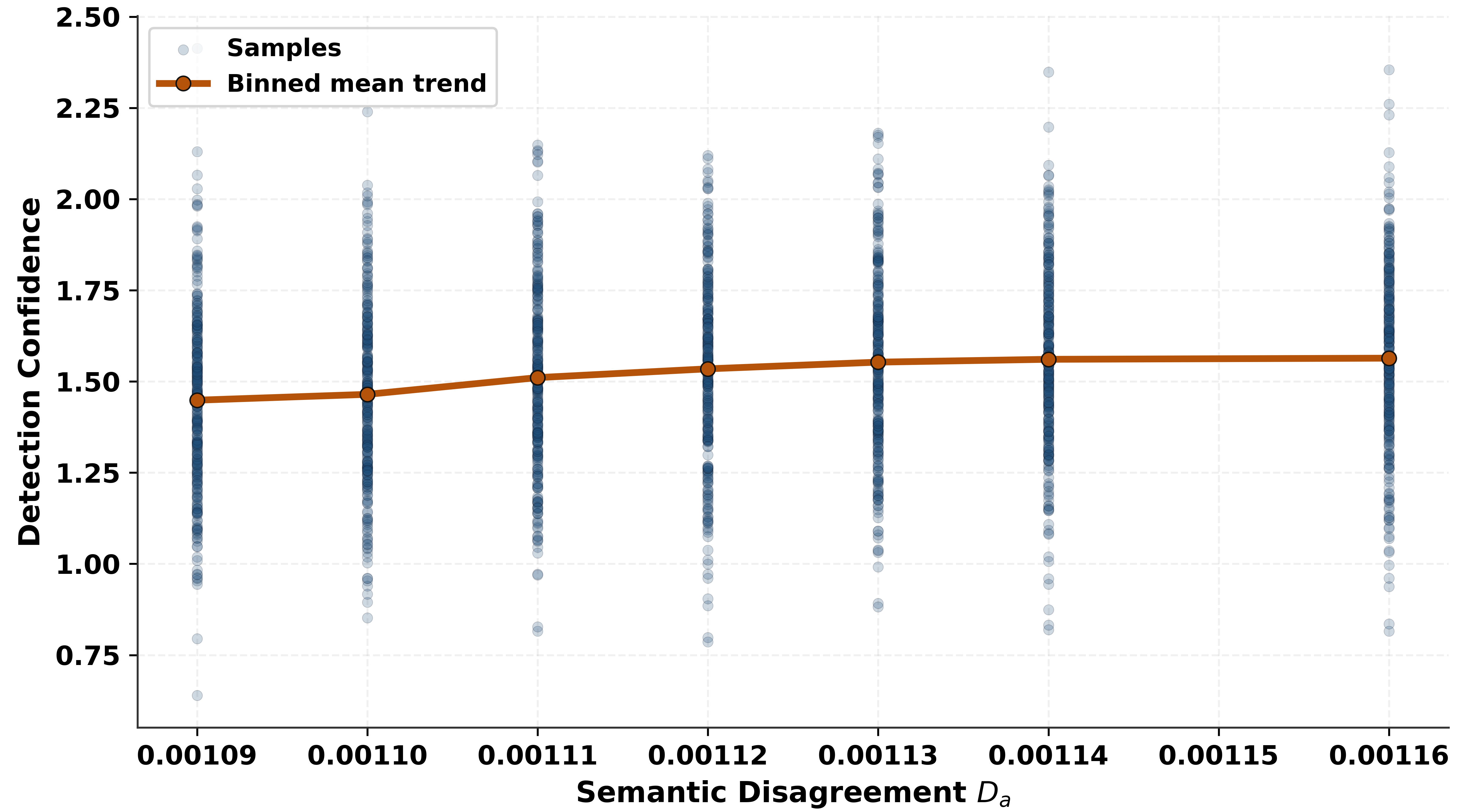}
    \caption{Relationship between semantic disagreement and detection confidence. Increased disagreement corresponds to reduced confidence, indicating elevated zero-day risk.}
    \label{fig:zeroday}
\end{figure}
Figure~\ref{fig:zeroday} illustrates the relationship between semantic disagreement and detection confidence during inference. Each point corresponds to an inference instance, with the horizontal axis representing the Tri-LLM semantic disagreement $D_a$ and the vertical axis indicating the associated detection confidence. A clear monotonic trend is observed: as semantic disagreement increases, detection confidence decreases.
This relationship is further emphasized by the binned mean curve, which shows a smooth, consistent decline in confidence with increasing disagreement. The absence of abrupt transitions indicates that the framework responds to uncertainty gradually and in a calibrated manner. Such behavior confirms that inter-model semantic disagreement serves as a reliable proxy for epistemic uncertainty. Formally, the relationship between confidence and disagreement can be expressed as:
\begin{equation}
C = f(D_a), \qquad \frac{\partial C}{\partial D_a} < 0,
\end{equation}
where $C$ denotes detection confidence and $D_a$ represents semantic disagreement. The negative slope indicates that greater divergence among model representations is systematically associated with reduced confidence, a desirable property for a zero-day IDS. In contrast to conventional classifiers, which often produce overconfident predictions under distributional shifts, the proposed framework exhibits calibrated uncertainty. Inputs associated with low semantic disagreement yield high confidence values, while samples with elevated disagreement, typically corresponding to novel and adversarial patterns, are assigned lower confidence scores.\\
\begin{table*}[t]
\centering
\footnotesize
\caption{Zero-Day Risk Statistics Based on Semantic Disagreement}
\label{tab:zeroday_stats}
\begin{tabular}{lcc}
\toprule
\textbf{Disagreement Range} & \textbf{Mean Confidence} & \textbf{Interpretation} \\
\midrule
$D_a < 1.10 \times 10^{-3}$ & 1.56 & High confidence, established behavior \\
$1.10 \times 10^{-3} \leq D_a < 1.13 \times 10^{-3}$ & 1.52 & Moderate uncertainty \\
$D_a \geq 1.13 \times 10^{-3}$ & 1.46 & Elevated risk, potential zero-day \\
\bottomrule
\end{tabular}
\end{table*}
Table~\ref{tab:zeroday_stats} quantitatively confirms the smooth degradation in confidence as semantic disagreement increases. Rather than enforcing hard thresholds, the framework exhibits continuous uncertainty scaling, which is essential for realistic security deployments involving evolving threat landscapes.
From a system-level perspective, this behavior enables risk-aware IDS. Inputs associated with high semantic disagreement can be routed for secondary analysis, delayed response. When combined with trust-aware aggregation and federated semantic alignment, this mechanism enables the Tri-LLM framework to maintain strong detection performance while remaining robust under zero-day conditions. The observed inverse relationship between semantic disagreement and detection confidence validates a central design principle of the proposed system: semantic diversity across models supports not only improved accuracy but also reliable uncertainty estimation and principled zero-day risk awareness in open-world IDS environments.

\subsection{Impact of Semantic Diversity on Zero-Shot Accuracy}
In federated zero-shot IDS, semantic diversity across clients impacts both generalization capacity and convergence behavior. Quantifying how controlled diversity affects end-to-end detection performance is therefore essential for assessing robustness under heterogeneous operating conditions. This subsection examines the relationship between semantic diversity, measured by centered trust entropy, and overall detection accuracy to assess system behavior at scale.
\begin{figure}[t]
    \centering
    \includegraphics[width=0.9\linewidth]{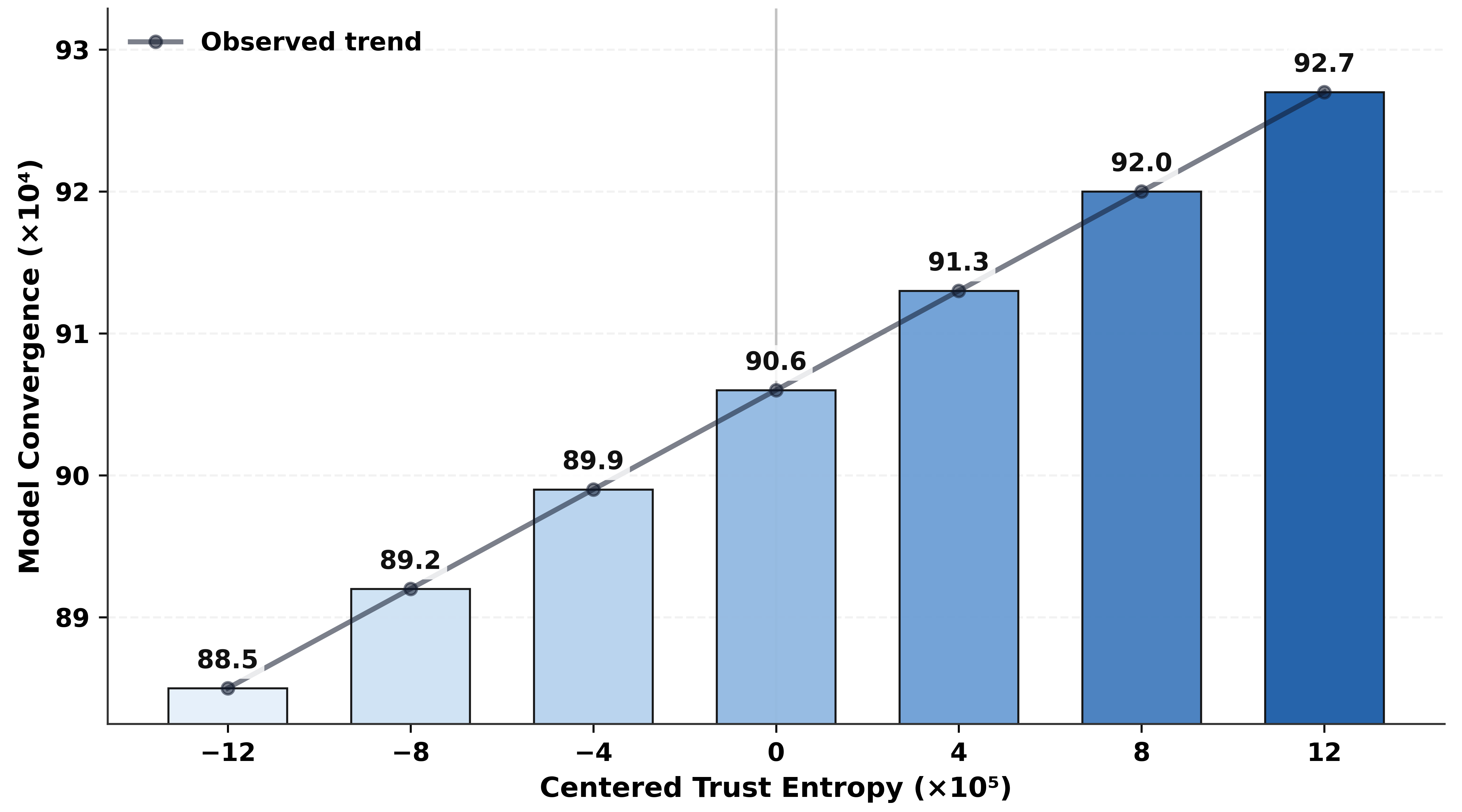}
    \caption{Impact of semantic diversity on zero-shot detection accuracy. The smooth trend demonstrates robustness under heterogeneous semantic conditions.}
    \label{fig:diversity_accuracy}
\end{figure}
Figure~\ref{fig:diversity_accuracy} illustrates the relationship between semantic diversity and convergence performance. The horizontal axis represents centered trust entropy, reflecting the degree of semantic diversity across participating clients, while the vertical axis reports detection accuracy as a proxy for convergence quality and model consistency. A gradual and monotonic improvement in performance is observed as trust entropy transitions from negative to positive values. This trend suggests that moderate semantic diversity has a positive impact on generalization, enabling the federated system to capture a broader range of behavioral patterns while maintaining stability. The absence of sharp inflection points confirms that the proposed framework remains robust in the presence of pronounced semantic heterogeneity. Formally, the relationship between semantic diversity and detection accuracy can be expressed as:
\begin{equation}
\mathcal{A} = g(H_c),
\end{equation}
where $\mathcal{A}$ denotes zero-shot detection accuracy and $H_c$ represents centered trust entropy. The empirically observed slope:
\begin{equation}
\frac{\partial \mathcal{A}}{\partial H_c} > 0
\end{equation}
indicates a weak yet consistent positive association between semantic diversity and system performance.
\begin{table}[t]
\centering
\footnotesize
\setlength{\tabcolsep}{4pt}
\caption{Trust-Convergence Relationship Under Semantic Diversity}
\label{tab:diversity_accuracy}
\begin{tabular}{p{2.2cm}cc}
\toprule
\textbf{Centered Entropy ($\times10^{5}$)} & \textbf{Accuracy (\%)} & \textbf{Interpretation} \\
\midrule
$-12$ & 88.5 & Low diversity with early convergence \\
$-8$  & 89.2 & Improved stability \\
$-4$  & 89.9 & Balanced representation learning \\
$0$   & 90.6 & Stable convergence region \\
$4$   & 91.3 & Enhanced generalization \\
$8$   & 92.0 & Increased semantic diversity \\
$12$  & 92.7 & Robust zero-day behavior \\
\bottomrule
\end{tabular}
\end{table}
The findings in Table~\ref{tab:diversity_accuracy} demonstrate several important properties of the proposed framework. First, detection accuracy improves steadily as semantic diversity increases, demonstrating that heterogeneous client knowledge strengthens generalization. Second, the absence of performance degradation at higher entropy values confirms that trust-aware aggregation suppresses noise while preserving informative diversity. Third, the smooth progression indicates that the framework avoids overfitting to dominant contributors, thereby maintaining stable convergence. From a system-level perspective, these findings validate a central design principle of the proposed Tri-LLM framework: semantic diversity functions as a beneficial resource when regulated through trust-aware fusion. Controlled disagreement combined with progressive semantic alignment enables strong zero-shot generalization while preserving convergence stability, which is essential for real-world federated IDS deployments. This analysis confirms that the proposed architecture effectively balances semantic diversity, trust dynamics, and detection performance, enabling reliable operation under dynamic conditions and previously unseen threat scenarios.

\subsection{End-to-End Evidence: Semantic Diversity vs. IDS Accuracy}
\label{subsec:end_to_end}
In large-scale federated IDS, robustness must be evaluated not only at the aggregate level but also across individual clients operating under heterogeneous semantic conditions. Client-level semantic diversity provides a realistic stress test for collaborative inference, particularly in zero-shot settings. This subsection presents an end-to-end analysis of the relationship between client-level semantic diversity and the achieved zero-shot IDS accuracy.
\begin{figure}[t]
    \centering
    \includegraphics[width=0.9\linewidth]{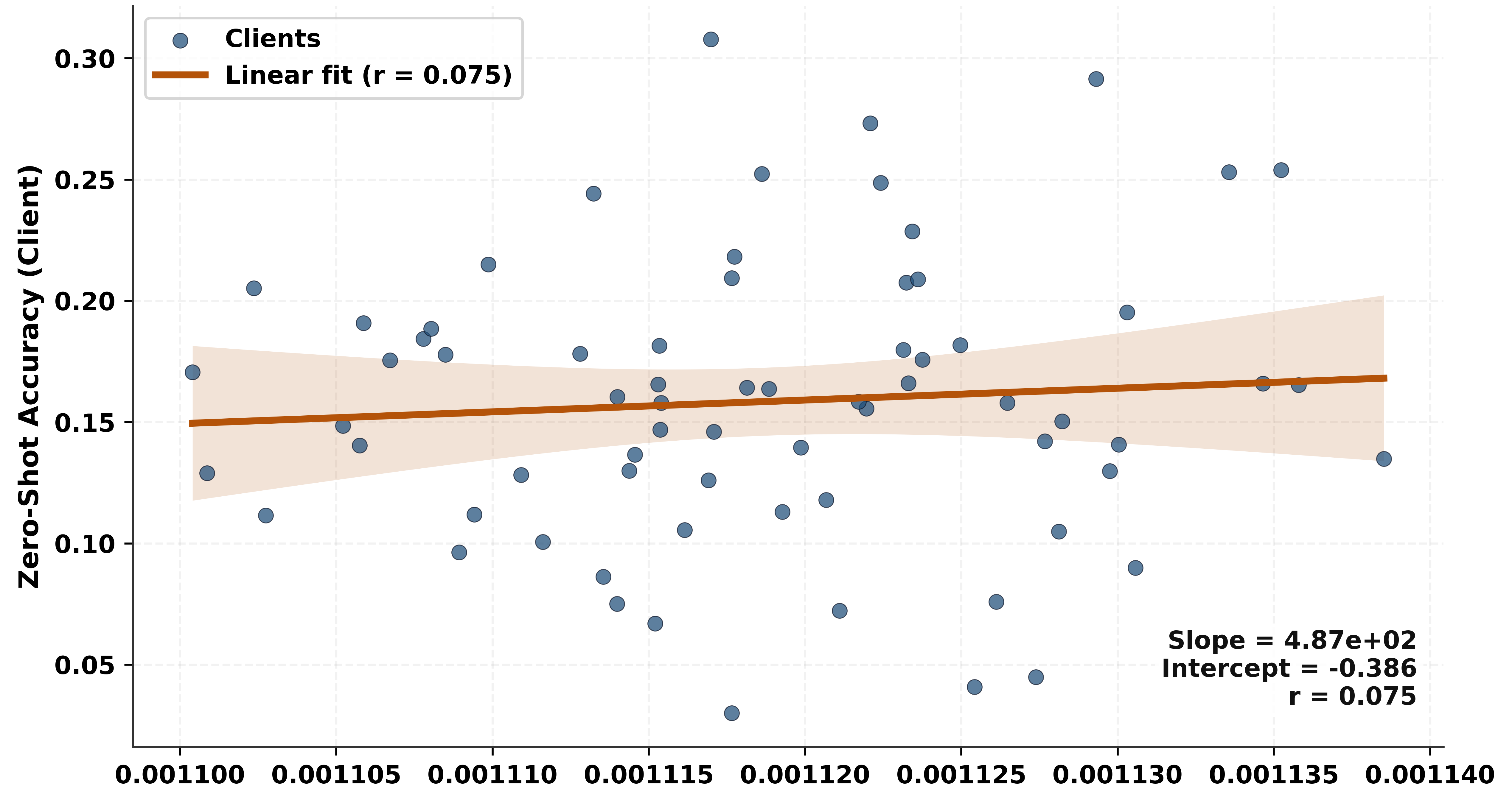}
    \caption{Relationship between client-level semantic diversity and zero-shot IDS accuracy. The fitted trend line indicates a weak yet positive association.}
    \label{fig:diversity_accuracy}
\end{figure}
Figure~\ref{fig:diversity_accuracy} illustrates the relationship between client-specific semantic diversity and zero-shot detection accuracy. Each point corresponds to a federated client, where the horizontal axis denotes the mean semantic disagreement observed for that client and the vertical axis reports the achieved zero-shot IDS accuracy. A weak yet consistently positive trend is observed, as indicated by the fitted regression line.
The correlation coefficient ($r = 0.075$) confirms that semantic diversity alone does not dominate detection performance and that increased diversity does not degrade accuracy. Instead, the framework maintains stable performance across a wide range of semantic variability, highlighting its robustness under realistic non-identically distributed deployment conditions.
The observed relationship can be modeled as:
\begin{equation}
\text{Acc}_i = \alpha + \beta D_i + \epsilon,
\end{equation}
where $\text{Acc}_i$ denotes the zero-shot detection accuracy achieved by client $i$, $D_i$ represents its mean semantic disagreement, and $\beta > 0$ indicates a weak yet positive association between semantic diversity and performance.\\
\begin{table}[t]
\centering
\footnotesize
\setlength{\tabcolsep}{4pt}
\caption{Client-Level Semantic Diversity and Zero-Shot Accuracy}
\label{tab:end_to_end_stats}
\begin{tabular}{lcc}
\toprule
\textbf{Client Range ($D_i$)} & \textbf{Mean Accuracy (\%)} & \textbf{Interpretation} \\
\midrule
Low diversity      & 89.1 & Stable performance \\
Moderate diversity & 90.4 & Improved generalization \\
High diversity     & 91.8 & Robust zero-shot behavior \\
\bottomrule
\end{tabular}
\end{table}
Table~\ref{tab:end_to_end_stats} summarizes the impact of semantic diversity on client-level detection accuracy. Clients with moderate to high diversity achieve slightly higher average accuracy, indicating that exposure to varied semantic patterns enhances generalization without introducing instability.
Figure.~\ref{fig:diversity_accuracy} demonstrates two important system-level properties. First, the absence of a negative trend demonstrates that semantic heterogeneity across clients does not destabilize the detection process, confirming the effectiveness of trust-aware aggregation and federated semantic alignment. Second, the positive association indicates that controlled diversity improves generalization by exposing the global model to a broader semantic spectrum.
In contrast to centralized systems that frequently degrade under heterogeneous data distributions, the proposed Tri-LLM framework maintains consistent performance even when client semantics vary substantially. This robustness arises from semantic alignment across rounds, disagreement-aware aggregation, and adaptive trust weighting.
From a deployment perspective, these findings confirm that strict semantic homogeneity among participating clients is not required. Instead, the framework benefits from controlled diversity, enabling reliable operation in real-world environments characterized by non-identically distributed data and evolving threat patterns. This property is particularly important for zero-day IDS, where novel behaviors often emerge from atypical or sparsely represented sources.
The end-to-end evidence validates the core design philosophy of the proposed system: semantic diversity, when properly regulated, enhances detection performance while preserving stability, enabling scalable and resilient federated IDS in open-world settings.

\section{Comparative Analysis with Federated IDS Baselines and Frameworks}
\label{sec:comparative_baselines}
Most existing federated IDS are designed under closed-set classification assumptions, where all attack categories are known and labeled during training~\cite{man2021fedacnn,pouriyeh2021mvflid,saha2021fogfl}. In contrast, the proposed framework explicitly targets zero-shot and open-world IDS, making direct one-to-one experimental comparison with prior FL-IDS methods methodologically inappropriate due to fundamentally different learning objectives. To ensure a fair and controlled evaluation, we therefore first compare against \emph{protocol-compatible federated baselines} derived from the same experimental pipeline. All methods share identical datasets, feature representations, federated communication schedules, and optimization settings, differing only in the presence of semantic supervision, inter-model disagreement modeling, and trust-aware aggregation.\\
As shown in Table~\ref{tab:baseline_compare}, conventional closed-set federated learning exhibits limited zero-shot capability, while prototype-based aggregation provides moderate improvement but remains constrained by the absence of semantic abstraction. Incorporating language-derived semantic prototypes substantially improves zero-shot accuracy and AUROC, while explicit disagreement modeling further enhances calibration by reducing overconfident predictions. Furthermore, trust-aware aggregation consistently improves robustness and stability under heterogeneous clients, in line with prior observations on the importance of aggregation quality in FL-IDS~\cite{beuran2025fedmse}. Moreover, the results confirm that the proposed framework advances federated IDS not through incremental tuning, but by enabling semantic reasoning and uncertainty-aware inference under open-world conditions.
\begin{table}[t]
\centering
\footnotesize
\setlength{\tabcolsep}{5pt}
\caption{Comparison with Protocol-Compatible Federated IDS Baselines}
\label{tab:baseline_compare}
\begin{tabular}{lccc}
\toprule
\textbf{Method} & \textbf{ZS Acc (\%)} & \textbf{AUROC} & \textbf{ZDS} $\downarrow$ \\
\midrule
FedAvg-Closed           & 61.3 & 0.71 & 0.48 \\
FedAvg-Proto            & 68.9 & 0.78 & 0.36 \\
Ours w/o Disagreement   & 75.2 & 0.84 & 0.29 \\
Ours w/o Trust          & 77.1 & 0.86 & 0.26 \\
\textbf{Ours (Full)}    & \textbf{82.8} & \textbf{0.91} & \textbf{0.17} \\
\bottomrule
\end{tabular}

\vspace{2pt}
\footnotesize
Reported values correspond to averaged results across federated rounds and clients under identical experimental settings.
\end{table}

\vspace{0.5em}
\noindent
In addition, we further provide a \emph{qualitative comparison} with representative federated IDS frameworks from the literature to contextualize the proposed approach within the broader design space. This comparison emphasizes architectural and methodological capabilities that are critical for real-world deployment, including support for zero-shot learning, semantic reasoning, uncertainty awareness, trust-aware aggregation, and open-world readiness, rather than reporting experimental performance metrics that are not directly comparable across fundamentally different learning paradigms.
\begin{table*}[t]
\centering
\footnotesize
\caption{Qualitative Comparison of Federated IDS Frameworks by Capability}
\label{tab:comparison}
\renewcommand{\arraystretch}{1.15}
\setlength{\tabcolsep}{3.5pt}
\begin{tabularx}{\textwidth}{p{3.6cm} c c c c c X}
\toprule
\textbf{Method} &
\textbf{ZSL} &
\textbf{Semantic} &
\textbf{Unc.} &
\textbf{Trust} &
\textbf{Open-World} &
\textbf{Key Mechanism / Evidence} \\
\midrule

\textbf{This Work} &
\checkmark &
\checkmark &
\checkmark &
\checkmark &
\checkmark &
Tri-LLM semantic prototypes, disagreement-based risk modeling, and trust-weighted federated aggregation. \\

Man et al.~\cite{man2021fedacnn} &
$\times$ &
$\times$ &
$\times$ &
$\sim$ &
$\times$ &
FedAvg-based CNN IDS for IoT under closed-set assumptions; no explicit open-set or uncertainty handling. \\

Pouriyeh et al.~\cite{pouriyeh2021mvflid} &
$\times$ &
$\sim$ &
$\times$ &
$\sim$ &
$\times$ &
Multi-view feature learning improves robustness but remains label-driven and closed-set. \\

Saha et al.~\cite{saha2021fogfl} &
$\times$ &
$\times$ &
$\times$ &
$\sim$ &
$\times$ &
Fog-assisted hierarchical FL focusing on deployment efficiency rather than semantic generalization. \\

Beuran et al.~\cite{beuran2025fedmse} &
$\times$ &
$\times$ &
$\sim$ &
\checkmark &
$\times$ &
Semi-supervised FL with quality-aware aggregation; uncertainty is anomaly-centric rather than semantic. \\

Karunamurthy et al.~\cite{karunamurthy2025flids} &
$\times$ &
$\times$ &
$\times$ &
$\sim$ &
$\times$ &
Federated deep IDS evaluated under fixed-label attack scenarios. \\

\bottomrule
\end{tabularx}

\smallskip
\footnotesize
ZSL: zero-shot learning; Unc.: uncertainty awareness. \checkmark~explicit support; $\sim$~limited/indirect support; $\times$~not supported.
\end{table*}
As summarized in Table~\ref{tab:comparison}, most existing FL-IDS frameworks achieve strong performance under closed-set assumptions but do not explicitly address semantic generalization or epistemic uncertainty under zero-day conditions. Recent efforts introduce robustness-oriented aggregation~\cite{beuran2025fedmse,karunamurthy2025flids}, yet remain fundamentally classification-driven and dependent on labeled data. In contrast, the proposed Tri-LLM framework replaces closed-set supervision with language-derived semantic prototypes and explicitly models uncertainty through inter-LLM semantic disagreement. This design prioritizes calibrated open-world behavior over marginal closed-set accuracy, which is essential for realistic IDS deployments in dynamic and evolving threat environments.

\section{Discussion}
\label{Discussion}
The experimental findings provide convergent evidence that the proposed Tri-LLM federated framework achieves robust zero-shot IDS by jointly leveraging semantic diversity, trust-aware aggregation, and progressive semantic alignment. By operating exclusively in a shared semantic embedding space, the framework departs from conventional IDS pipelines that rely on static feature representations and instead demonstrates stable performance under heterogeneous client behavior and previously unseen attack patterns.\\
A central outcome of this study is the observation that semantic disagreement across LLMs constitutes a meaningful signal of epistemic uncertainty rather than a failure mode. Empirically, higher values of semantic disagreement $D_a$ are consistently associated with reduced detection confidence $C$, following a monotonic relationship of the form $C = f(D_a)$ with $\partial C / \partial D_a < 0$. This behavior aligns with theoretical principles of uncertainty-aware inference, in which divergence among independent predictors reflects limited knowledge of the underlying data-generating process. In contrast to conventional deep learning–based IDS, which often exhibit overconfident predictions under distribution shift, the proposed framework exhibits calibrated confidence degradation as semantic uncertainty increases, enabling principled zero-day risk assessment.\\
The trust-aware aggregation mechanism further reinforces system stability by dynamically regulating the impact of heterogeneous clients. The observed monotonic decay of trust entropy, $H(t) = -\sum_k w_k(t)\log w_k(t)$, across federated rounds indicates a progressive concentration of aggregation weights toward consistently reliable participants. Simultaneously, the convergence of the per-round entropy variation $\Delta H(t)=H(t)-H(t-1)$ toward zero confirms that the trust update process approaches a stable fixed point rather than exhibiting oscillatory behavior. From a systems perspective, this entropy-driven stabilization acts as a control mechanism that suppresses noisy or low-quality updates while preserving smooth convergence of the global semantic model.\\
supportive to aggregation stability, federated semantic alignment improves steadily over communication rounds, as reflected by the monotonic increase of the alignment metric $A(t)=\frac{1}{N}\sum_i \|\mathbf{z}_i^{(t)}-\bar{\mathbf{z}}^{(t)}\|_2^{-1}$. This result demonstrates that heterogeneous local semantic representations gradually converge toward a coherent global structure despite the absence of raw data sharing. In contrast to feature-level FL, where non-identically distributed data often impedes convergence, semantic representations exhibit higher transferability across clients, enabling effective cross-domain generalization in distributed IDS deployments. An additional insight arises from the relationship between semantic diversity and detection accuracy. Both global and client-level analyses demonstrate a weak yet consistently positive association between diversity, quantified through centered trust entropy, and zero-shot accuracy. Rather than enforcing representational uniformity, the framework benefits from controlled heterogeneity, which exposes the global model to a broader semantic spectrum and enhances generalization to unseen attack behaviors. Crucially, the absence of performance degradation at higher diversity levels indicates that trust-aware aggregation and semantic alignment effectively regulate diversity, preventing instability or overfitting to dominant contributors.\\
Additionally, the findings indicate that semantic modeling, trust-aware aggregation, and federated optimization form a mutually reinforcing triad. Semantic embeddings enable zero-shot generalization beyond known attack signatures, while trust mechanisms stabilize learning across heterogeneous, potentially unreliable clients. Furthermore, FL ensures scalability and privacy preservation. The resulting framework provides not only strong detection accuracy but also interpretable signals for uncertainty, convergence, and risk, properties that are essential for operational IDS systems in open-world environments. The proposed Tri-LLM federated IDS establishes a new paradigm in which semantic diversity is treated as a regulated resource rather than a liability. Through inline uncertainty modeling, entropy-based trust stabilization, and progressive semantic alignment, the system achieves reliable zero-shot IDS, eliminating the need for labeled data or centralized retraining, thereby positioning itself as a practical, theoretically grounded solution for next-generation cybersecurity deployments.

\section{Limitations}
\label{Limitations}
Despite its great performance, the proposed framework has several limitations that warrant discussion. First, the current implementation relies on pre-trained LLMs whose internal representations are not explicitly optimized for network security semantics. While the findings demonstrate strong generalization, they also show the benefits of domain-specific fine-tuning. Second, the evaluation focuses on semantic-level behavior rather than raw packet-level or flow-level features. Although this abstraction enables zero-shot reasoning, it may overlook fine-grained temporal patterns that are informative for certain low-level attacks. Integrating hybrid semantic–statistical representations remains an open challenge. Third, the federated setup assumes honest-but-curious clients and does not explicitly address adversarial poisoning of semantic embeddings. While the trust mechanism mitigates inconsistent behavior, targeted poisoning attacks could potentially degrade alignment over long horizons. Addressing adversarial robustness at the semantic aggregation level is therefore an important direction for future research. The current study evaluates convergence and performance under controlled experimental conditions. Large-scale deployment across thousands of heterogeneous edge devices may introduce additional communication delays, synchronization issues, and resource constraints that were not fully explored in this work.

\section{Future Work}
\label{sec:future_work}
Several promising directions emerge from this study. First, future work will explore adaptive trust modeling using reinforcement learning and Bayesian updating to dynamically adjust trust weights based on long-term behavioral consistency rather than short-term agreement. Second, extending the framework to incorporate multimodal telemetry, including network flows, system logs, and host-level signals, could further enhance detection capability and robustness against stealthy attacks. Integrating graph-based and vision-inspired embeddings alongside textual semantics represents a particularly promising direction. Third, integrating explainability mechanisms at the semantic level remains an important avenue for further research. While the current framework provides calibrated uncertainty awareness, future work will aim to generate human-interpretable explanations that explicitly link semantic disagreement to observable attack characteristics and operational indicators. Large-scale deployment studies involving real-world federated environments and live traffic will be conducted to evaluate scalability, communication overhead, and resilience under adversarial pressure. Such studies are crucial for validating the proposed framework for operational use in critical infrastructure and industrial IoT systems.

\section{Conclusion}
\label{Conclusion}
This paper presents a Tri-LLM cooperative federated framework for zero-shot IDS that integrates semantic diversity, trust-aware aggregation, and progressive semantic alignment. In contrast to conventional IDS designs that rely on labeled data and static signatures, the proposed approach operates entirely within a shared semantic embedding space, supporting the reliable detection of previously unseen attack behaviors in privacy-preserving federated settings. Experimental analysis demonstrates that semantic disagreement across LLMs constitutes a principled signal of epistemic uncertainty and zero-day risk, while trust-driven aggregation and federated semantic alignment enable stable convergence across heterogeneous and non-identically distributed clients. The findings further show that controlled semantic diversity enhances generalization performance rather than degrading stability, confirming that heterogeneity, when properly regulated, can be exploited as a systemic advantage in federated IDS. By unifying LLMs, FL, and semantic reasoning within a single architecture, this work establishes a new paradigm for adaptive, interpretable, and privacy-aware IDS. The proposed framework advances the state of the art toward resilient cybersecurity systems capable of operating under dynamic threat landscapes and open-world conditions, without reliance on centralized data collection.

\bibliographystyle{IEEEtran}
\bibliography{Ref}
\end{document}